%% file: Strucutre-Thms.tex
\documentclass[12pt,onecolumn]{IEEETran}
\pdfoutput=1 
\usepackage{amssymb, amsmath, amsthm }
\usepackage{graphicx,subfig}
\usepackage{setspace, bbm}

\input{Defs}

\begin{document}
\title{Structure Theorems for Real--Time Variable Rate Coding With and Without Side Information}
\author{Yonatan Kaspi and Neri Merhav\\

Department of Electrical Engineering \\
Technion - Israel Institute of Technology \\
Technion City, Haifa 32000, Israel\\
Email: \{kaspi@tx, merhav@ee\}.technion.ac.il}

\maketitle

%\begin{center}
%Department of Electrical Engineering \\
%Technion - Israel Institute of Technology \\
%Haifa 32000, ISRAEL \\
%Email: kaspi@tx.technion.ac.il
%\end{center}

%\vspace{1.5\baselineskip}
%\setlength{\baselineskip}{1.5\baselineskip}

\begin{abstract}
The output of a discrete Markov source is to be encoded instantaneously by a variable--rate encoder and decoded by a finite--state decoder. Our performance measure is a linear combination of the distortion and the instantaneous rate. Structure theorems, pertaining to the encoder and next--state functions are derived for every given finite--state decoder, which can have access to side information.
\end{abstract}

\input{Introduction}
\input{prelims}

%\input{MainResults}
\input{NoSI}
\input{MarkovMem}

\section{Side Information at the Decoder}\label{Sec:SI}
\subsection{Preliminaries and main result}
\begin{figure}[htp]
\centering
\includegraphics[width=0.8\textwidth]{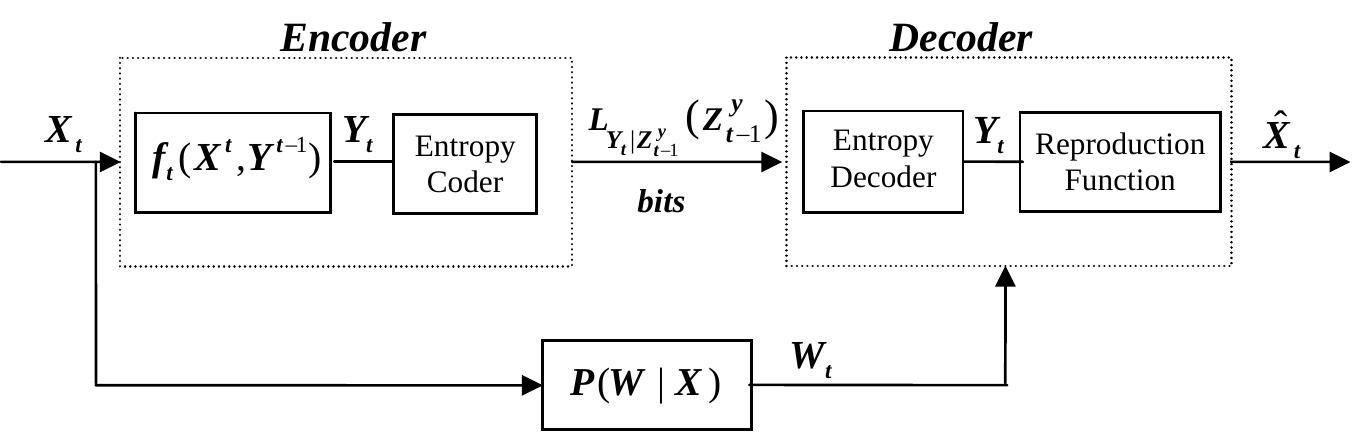}
\caption{System model with SI.\label{Fig:ModelSI}}
\end{figure}

In this section, we assume that the decoder has access to SI. The SI sequence, $W_1,W_2,...,W_T$, $W_t\in \calW$, is generated by a discrete memoryless channel (DMC), fed by $X_1,X_2,\ldots,X_T$:
\begin{align}
    P(w_1,\ldots,w_T|x_1,\ldots x_T)=\prod_{t=1}^T P(w_t|x_t).\nt
\end{align}
For simplicity, we assume that $P(w|x)>0$ for all $X\in\calX$ and $W\in\calW$. Our results, however, will continue to hold without this assumption with minor changes to the length function (see \cite{AlonOrlitski96}).
The SI is used both in the reproduction function and in the state update function. We assume that the state now consists of two sub--states. The first, $Z_t^y\in\calZ^y$, is independent of the SI and is updated as in Section \ref{Sec:Prelim}. The second, $Z_t^w\in\calZ^w$, is updated by
\begin{align}
    Z^w_1 &= r^w_1(W_1,Y_1),\nt\\
    Z^w_t &= r^w_t(W_t, Y_t, Z^w_{t-1}), ~~~~ t=2,3,\ldots,T.
\end{align}
The reproduction symbols are produced by a sequence of functions $\left\{g_t\right\}$, $g_t:\calY\times\calW\times\calZ^w\times\calZ^y\to\hat{\calX}$ as follows:
\begin{align}
    \hX_1 &= g_1(W_1,Y_1),\nt\\
    \hX_t &= g_t(W_t, Y_t,Z_{t-1}^w, Z_{t-1}^y), ~~~~ t=2,3,\ldots,T.
\end{align}
Since $Z_t^w$ is not known at the encoder, it cannot be used by the variable--length encoder and thus the cost function is now given by
\begin{align}
    J_t = \bE\left\{\rho_t(X_t, g_t(W_t,Y_t,Z_{t-1}^w, Z_{t-1}^y)) +\lambda L_{Y_t}(Z_{t-1}^y) \right\}
\end{align}
Let $B_t\eqd P_{Z_t^w|X^t}(\cdot|X^t)$ and $b_t \eqd P_{Z_t^W|X^t}(\cdot|x^t)$, i.e., $b_t\in\mathbb{R}^{\calZ^w}$ is a probability measure over the sub--state of the decoder, $Z_t^w$, which is not known to the encoder. Note that since the decoder does not have access to $x^t$, $b_t$ is not known to the decoder.
Our system model with SI is depicted in Figure \ref{Fig:ModelSI}.

The following two theorems are the contribution of this section.
\begin{theorem}\label{Thm:SI}
    For a Markov source and any given sequence of memory update functions $\{r_t\}$, reproduction functions $\{g_t\}$ and distortion measures $\{\rho_t\}$, there exists a sequence of deterministic encoders $Y_t = f_t(B_{t-1},X_t,Z_{t-1}^y)$, which is optimal.
\end{theorem}
The last theorem basically states that the results of \cite{Teneketzis2006} continue to hold in this setting as well.

As in Section \ref{Sec:NoSI}, When $Z_t^y=Y^t$, we have the counterpart of Theorem \ref{Thm:InfMem} for the SI setting when the optimal reproduction functions are used:
\begin{theorem}\label{Thm:InfMemSI}
    For a Markov source and any given sequence of SI memory update functions $\{r_t^w\}$ and distortion measures $\{\rho_t\}$, when $Z_t^y=Y^t$ and the optimal reproduction function are used, there exists a sequence of deterministic encoders $Y_t=f_t(P_{X_t,Z_{t-1}^w|Y_{t-1}}(\cdot,\cdot|y^{t-1}),X_t)$ which is optimal.
\end{theorem}

Note that unlike the result of Theorem \ref{Thm:SI}, in the setting of Theorem \ref{Thm:InfMemSI}, the encoder does not need to store $B_{t-1}$ which is a function of $X^{t-1}$. Instead it stores the joint conditional probability measure of $(X_t,Z_{t-1}^w)$, which is a function of $Y^{t-1}$. There is no contradiction between the theorems since the setting of Theorem \ref{Thm:InfMemSI} is different both in the use of the optimal reproduction functions and in the SI independent sub--state of the decoder.

The proof of Theorem \ref{Thm:SI} follows the lines of the proof of Theorem \ref{Thm:NoSI} after Lemmas \ref{Lem:2Stage}-\ref{Lem:3StageConcave} are extended to the setting of this section.
The changes to Lemmas \ref{Lem:2StageConcave}, \ref{Lem:3StageConcave} are quite simple (roughly speaking, instead of $x_t$ write $(b_{t-1},x_t)$ everywhere in the proof). The extension of the two-- and three--stage lemmas (Lemmas \ref{Lem:2Stage},\ref{Lem:3Stage}) is more involved and is given in the next two subsections. After these lemmas will be proven, the remainder of the proof is the same as in the previous section and therefore, will be omitted. Theorem \ref{Thm:InfMemSI} is proved in Subsection \ref{Sec:InfMemSI}.

\subsection{Theorem \ref{Thm:SI} proof outline}
We redefine $B_t,b_t$ to be $B_t\eqd P_{Z_t^w|X^t,Y^t}(\cdot|X^t,Y^t)$ and $b_t \eqd P_{Z_t^W|X^t,Y^t}(\cdot|x^t,y^t)$. Since Theorem \ref{Thm:SI} states that the encoders can be deterministic, the conditioning on $Y^t$ in the definition of $B_t$ is redundant since the sequence of encoder outputs $Y^t$ is a deterministic function of the source symbols $X^t$. However, in the proof of Theorem \ref{Thm:SI}, since we are allowing stochastic encoders a-priori, $Y^t$ adds information to $X^t$ and therefore, this conditioning is needed.
We precede the proof of this theorem with a short discussion regarding its significance. Since $b_{t-1}$ is a deterministic function of $(x^{t-1},y^{t-1})$, one may argue that this theorem does not simplify the structure of the general encoder, which is, anyway, a function of $x^t,y^{t-1}$. However, it turns out that the encoder can update $b_t$ recursively using only the data that is available to it at each stage (i.e., $X_t,Y_t,B_{t-1}$). To see why this is true, observe that
\begin{align}
    b_t(z)=P(z_t^w=z|x^t,y^t) &= P(r_t^w(y_t,w_t,z_{t-1}^w)=z|x^t,y^t)\nt\\
    &=\sum_{w_t,z_{t-1}^w : r_t^w(y_t,w_t,z_{t-1}^w)=z} P(w_t,z_{t-1}^w|x^t,y^t)\nt\\
    &=\sum_{w_t,z_{t-1}^w : r_t^w(y_t,w_t,z_{t-1}^w)=z} P(w_t|x^t,y^t)P(z_{t-1}^w|w_t,x^t,y^t)\nt\\
    &=\sum_{w_t,z_{t-1}^w : r_t^w(y_t,w_t,z_{t-1}^w)=z} P(w_t|x_t)P(z_{t-1}^w|x^{t-1},y^{t-1})\nt\\
    &\eqd h(b_{t-1},x_t,y_t,z)\label{eq:BRecursive}
\end{align}
Since this is true for any $z\in\calZ^w$, we showed that $b_t$ is a function of $(b_{t-1},x_t,y_t)$. Therefore, the encoder can recursively update $b_t$ at the end of each encoding stage using its knowledge of $(b_{t-1}, x_t)$ and its last output $y_t$.

\subsubsection{Two-stage lemma}\label{Sec:2StageLemmaSI}
We start by analyzing a system with only two stages, where the first encoder is known.
\begin{lemma}\label{Lem:2StageSI}
    In a two--stage system ($T=2$), there exists a deterministic second--stage encoder, $Y_2=f_2(B_1,X_2,Z_1^y)$ which is optimal.
\end{lemma}
\textit{Proof of Lemma \ref{Lem:2StageSI}: }
Note that $J_1$ is unchanged by changing the second stage encoder. Denote the set of stochastic encoders which are functions of $(X_1,X_2,Y_1)$ by $\{f_{X^2Y_1}^s \}$. The minimization of $J_2$ can be written as
\begin{align}
    &J_2^* =
    %&=\min_{f}\bE\left\{\bE\left[\rho_2(X_2, g_t(W_2,Y_2,Z_{1}^w, Z_{1}^y)) +L_{Y_2|Z_1^y}(Z_{1}^y) \bigg| X_1,X_2,Y_1,Y_2\right]\right\}\nt\\
    \inf_{\{f_{X^2Y_1}^s \}}\bE\left\{\bE\left[\rho_2(X_2, g_t(W_2,Y_2,Z_{1}^w, z_{1}^y))+\lambda L_{Y_2|Z_1^y}(Z_{1}^y) | X_1,X_2,Y_1,Y_2,Z_1^y\right]\right\}\nt\\
    &=\inf_{\{f_{X^2Y_1}^s \}}\bE\left\{\lambda L_{Y_2|Z_1^y}(Z_{1}^y)+\bE\left[\rho_2(X_2, g_t(W_2,Y_2,Z_{1}^w, Z_{1}^y))| X_1,X_2,Y_1,Y_2,Z_1^y\right]\right\}.\label{eq:J2SI}
\end{align}
\normalsize
Focusing on the inner conditional expectation, we have
\begin{align}
    &\bE\left[\rho_2(X_2, g_t(W_2,Y_2,Z_{1}^w, Z_{1}^y))| X_1,X_2,Y_1,Y_2,Z_1^y\right]=\nt\\
    %&=\sum_{w_2,z_1^w}P(w_2,z_1^w|X_1,X_2,Y_1,Y_2,Z_1^y)\rho_2(X_2, g_t(w_2,Y_2,z_{1}^w, Z_{1}^y))\nt\\
    &=\sum_{w_2,z_1^w}P(w_2|X_2)P(z_1^w|X_1,Y_1)\rho_2(X_2, g_t(w_2,Y_2,z_{1}^w, Z_{1}^y))\nt\\
    %&=\min_{f(X_1,X_2)}\bE\left\{\sum_{z_1^w}P(z_1^w|x_1)\tilde{\rho}_2(X_2,y_2,z_{1}^w, z_{1}^y) +L_{Y_2|Z_1^y}(Z_{1}^y) \nt\\
    &\eqd\hat{\rho}_2(B_1,X_2,Y_2,Z_{1}^y) \label{eq:ModDist}
\end{align}
where $B_1\eqd P_{z_1^w|x_1,y_1}(\cdot|X_1,Y_1)$ is a probability measure on $Z_1^w$ that represents the encoder's belief on the decoder's unknown state. Note that $B_1$ is a deterministic function of $(X_1,Y_1)$ and the modified distortion measure \eqref{eq:ModDist} depends on $(X_1,Y_1)$ only through $B_1$.
Combining \eqref{eq:J2SI} and \eqref{eq:ModDist} we have
\begin{align}
    &J_2^* = \inf_{\{f_{X^2Y_1}^s \}}\bE\left\{\hat{\rho}_2(X_2,Y_2,B_1, Z_{1}^y)+\lambda L_{Y_2|Z_1^y}(Z_{1}^y)\right\}\label{eq:J2SiOpt}
\end{align}
where the expectation is with respect to $P(b_1,x_2,y_2,z_1^y)$.
Consider the quadruple of RV's $(B_1,X_2,Y_2,Z_1^y)$. We have
\begin{align}
    P(b_1,x_2,y_2,z_1^y)=P(b_1,x_2,z_1^y)P(y_2|b_1,x_2,z_1^y).
\end{align}
While $P(b_1,x_2,z_1^y)$, which depends on the first stage design and the source, remains fixed in the optimization in \eqref{eq:J2SiOpt} (it can be thought of a state of the system, governed by the choice of the first stage design),  $P(y_2|b_1,x_2,z_1^y)$ depends on the second stage encoder since
\begin{align}
    &P(y_2|b_1,x_2,z_1^y)  =\nt\\
%    &=\frac{\sum_{x_1,y_1}P(x_1,x_2,y_1)P(y_2|x_1,x_2,y_1)P(b_1|x_1,x_2,y_1,y_2)P(z_1^y|b_1,x_1,y_1,x_2,y_2)}
 %   {P(b_1,x_2,z_1^y)}\nt\\
    & \frac {\sum_{x_1,y_1}P(x_1,x_2,y_1)P(y_2|x_1,x_2,y_1)P(b_1|x_1,y_1)P(z_1^y|y_1)}{P(b_1,x_2,z_1^y)}\label{eq:beyes2}
\end{align}
in the last expression, $P(b_1|x_1,y_1)=1$ for all $x_1,y_1$ that yield the same specific conditional distribution, $b_1$, over $Z_1^w$ and zero otherwise. $P(y_2|x_1,x_2,y_1)$ is governed by the second stage stochastic encoder, which maps $(x_1,x_2,y_1)$ to a probability measure on $\calY$. Let us now look at the expectation in \eqref{eq:J2SiOpt}:
\begin{align}
    &\bE\left\{\hat{\rho}_2(X_2,Y_2,B_1, Z_{1}^y) +L_{Y_2|Z_1^y}(Z_{1}^y) \right\} =\nt\\
    &\sum_{b_1,x_2,y_2,z_1^y}P(b_1,x_2,z_1^y)P(y_2|b_1,x_2,z_1^y)\Bigg\{\hat{\rho}_2(x_2,y_2,b_1, z_{1}^y) +\nt\\
     &~~~ \min_{l(\cdot)\in\calA}\sum_{b_1',x_2',y_2'}P(y_2'|b_1',x_2',z_1^y)P(b_1',x_2'|z_1^y)l(y_2')\Bigg\}.\label{eq:E}
\end{align}
\normalsize
As in the proof of Lemma \ref{Lem:2Stage}, from \eqref{eq:E} we see that the optimization will be affected by the choice of the second stage encoder through $P(y_2'|b_1',x_2',z_1^y)$ for all $(b_1',x_2',z_1^y)$. Denote the subset of stochastic second stage encoders that are functions of $(b_1,x_2,z_1^y)$ by $\{f_{B_1X_2Z_1^y}^s\}$. Since $(b_1,z_1^y)$ are functions of $(x_1,y_1)$, $\{f_{B_1X_2Z_1^y}^s\} \subset \{f_{X_1X_2Y_1}^s\}$. From \eqref{eq:beyes2} we see that every specific $f_2\in f_{X_1X_2Y_1}^s$ is mapped to some specific $\hat{f}_2\in f_{B_1X_2Z_1^y}^s$. Since the optimization is affected only by $P(y_2|b_1,x_2,z_1^y)$, if instead of using a specific $f_2$, we would use $\hat{f}_2$ that result from it, we would not change the joint probability of the quadruple $(B_1,X_2,Y_2,Z_1^y)$ and thus the second stage cost will not be changed. Therefore we conclude that
\begin{align}
    \inf_{\{f_{X_1X_2Y_1}^s\}}J_2=\inf_{\{f_{B_1X_2Z_1^y}^s\}}J_2.
\end{align}
To complete the proof, we need to show that it is enough to search in the finite subset of deterministic functions of $(b_1,x_2,z_1^y)$, which we denote by $\{f_{B_1X_2Z_1^y}^d\}$. This is done by repeating the arguments we used below \eqref{eq:AlternativeMin} in the end of the proof of Lemma \ref{Lem:2Stage}.
\IEEEQED\\

\subsubsection{Three-stage Lemma}
\begin{lemma}\label{Lem:3StageSI}
    In a three-stage system $(T=3)$ with a Markov source, if the third--stage encoder is a deterministic function of $(B_2,X_3,Z_2^y)$, then there exists a deterministic second stage encoder $Y_2=f_2(B_1,X_2,Z_1^y)$ which is optimal.
\end{lemma}
\textit{Proof of Lemma \ref{Lem:3StageSI}: }
We define, as in we did in Subsection \ref{Sec:2StageLemmaSI}, $\{f_{X_1X_2Y_1}^s\}$ to be the set of all possible stochastic second stage encoders. Let $\{f_{B_1X_2Z_1^y}^s\}\subset\{f_{X_1X_2Y_1}^s\}$ be the subset that contains all stochastic second stage encoders that are functions of $(B_1,X_2,Z_1^y)$ and finally, let $\{f_{B_1X_2Z_1^y}^d\}\subset\{f_{B_1X_2Z_1^y}^s\}$ denote the set of deterministic encoders which are functions of $(B_1,X_2,Z_1^y)$.
Since the first stage is fixed, $J_1$ is unaffected. Our goal is to jointly optimize $(J_2+J_3)$ with respect to the second stage encoder and show that
\begin{align}
    \inf_{\{f_{X_1X_2Y_1}^s\}}\lb J_2+J_3\rb = \min_{\{f_{B_1X_2Z_1^y}^d\}}\lb J_2+J_3 \rb.\label{eq:3StageGoal}
\end{align}
We start by focusing on the third stage cost.
\begin{align}
    &J_3 = \bE\left\{\rho_3(X_3,g_3(Y_3,W_3,Z_2^w,Z_2^y))+ \lambda L_{Y_3|Z_2^y}(Z_2^y)\right\}\nt\\
    & = \bE\bigg\{\lambda L_{Y_3|Z_2^y}(Z_2^y)+\bE\left\{\rho_3(X_3,g_3(Y_3,W_3,Z_2^w,Z_2^y)) \bigg| X^3,Y^2,Z_2^y\right\}\bigg\} \label{eq:J_3}
\end{align}
\normalsize
Focusing on the inner expectation of \eqref{eq:J_3}, we have
\begin{align}
    &\bE\left\{\rho_3(X_3,g_3(Y_3,W_3,Z_2^w,Z_2^y)) \Bigg| X^3,Y^2,Z_2^y\right\}\nt\\
    &= \bE\left\{\rho_3(X_3,g_3(Y_3,W_3,Z_2^w,Z_2^y)) \Bigg| X^3,Y^3,Z_2^y\right\}\nt\\
    &= \sum_{w_3,z_2^w}P(w_3|X_3)P(z_2^w|X^2,Y^2)\rho_3(X_3,g_3(Y_3,w_3,z_2^w,Z_2^y))\nt\\
    &\eqd \hat{\rho}_3(B_2,X_3,Y_3,Z_2^y)
\end{align}
\normalsize
where the first equality is true since $Y_3$ is a function of $(B_2,X_3,Z_2^y)$ and $B_2$ is a deterministic function of $(X^2,Y^2)$. Therefore,
\begin{align}
    J_3 &= \sum_{b_2,x_3,y_3,z_2^y}P(b_2,x_3,z_2^y)P(y_3|b_2,x_3,z_2^y)\times\nt\\
     &~~~~\left[\hat{\rho}_3(b_2,x_3,y_3,z_2^y) + \lambda L_{Y_3|Z_2^y}(z_2^y)\right].\label{eq:J3SI}
\end{align}
In the last expression, $P(y_3|b_2,x_3,z_2^y)$ will not be affected by the optimization of the second stage encoder since, under the assumptions of Lemma \ref{Lem:3StageSI}, the third encoder is a fixed deterministic function of $(b_2,x_3,z_2^y)$ (i.e., $P(y_3|b_2,x_3,z_2^y)=\ind{f_3(b_2,x_3,z_2^y)=y_3})$. Thus, the second stage encoder affects the last expression only through $P(b_2,x_3,z_2^y)$ since
\begin{align}
    &P(b_2,x_3,z_2^y) = \sum_{b_1,x_2,y_2}P(b_1,x_2,z_1^y)P(y_2|b_1,x_2,z_1^y)\times\nt\\
    &~~~~~~~~P(x_3|x_2)\ind{h(b_1,x_2,y_2)=b_2}\ind{r^y_2(z_1^y,y_2)=z_2^y}\label{eq:beyes3},
\end{align}
where $h(b_1,x_2,y_2)$ was defined in \eqref{eq:BRecursive} and we used the fact that the source is Markov. As in subsection \ref{Sec:2StageLemmaSI}, $P(b_1,x_2,z_1^y)$ is the result of the first stage design and the source. Note that $\ind{h(b_1,x_2,y_2)=b_2}$, $\ind{r^y_2(z_1^y,y_2)=z_2^y}$ are not affected by the choice of the second stage encoder since they represent known deterministic functions of $(b_1,x_2,y_2)$ and $(z_1^y,y_2)$ respectively.
Focusing on the third stage average codeword length for $Z_2^y=z_2^y$, we have
\begin{align}
    &L_{Y_3|Z_2^y}(z_2^y) = \min_{l(\cdot)\in \calA} \sum_{b_2,x_3,y_3}P(b_2,x_3,y_3|z_2^y)l(y_3)\nt\\
    &= \min_{l(\cdot)\in \calA} \sum_{b_2,x_3,y_3}P(b_2,x_3|z_2^y)P(y_3|b_2,x_3,z_2^y)l(y_3)\label{eq:L3}.
\end{align}
\normalsize
Again, the second--stage encoder affects only $P(b_2,x_3,z_2^y)$ (and thus $P(b_2,x_3|z_2^y)$).
In \eqref{eq:beyes2},\eqref{eq:E} we showed that the second--stage encoder affect the second stage cost only through $P(y_2|b_1,x_2,z_1^y)$. In \eqref{eq:J3SI},\eqref{eq:beyes3},\eqref{eq:L3} we showed that the third stage cost depends on the second stage cost only through $P(y_2|b_1,x_2,z_1^y)$. Therefore, we conclude that the optimization of the second stage encoder affects $(J_2+J_3)$ only through $P(y_2|b_1,x_2,z_1^y)$. Repeating the arguments we used in the proof of Lemma \ref{Lem:2StageSI}, if we use $\hat{f}_2\in f_{B_1X_2Z_1^y}^s$ that result from a specific $f_2\in f_{X_1X_2Y_1}^s$ (through \eqref{eq:beyes2}) instead of using that specific $f_2$, we would not change the joint probability of $(B^2,X^3,Y^3,Z_1^y,Z_2^y)$ and therefore will not change the value of $(J_2+J_3)$. Therefore we have
\begin{align}
    \min_{\{f_{X_1X_2Y_1}^s\}}\lb J_2+J_3\rb = \min_{\{f_{B_1X_2Z_1^y}^s\}}\lb J_2+J_3 \rb.
\end{align}
From here, the same arguments we used after \eqref{eq:AlmostDone} in the proof of Lemma \ref{Lem:3Stage} will complete the proof.

\subsection{Infinite memory decoder - proof of Theorem \ref{Thm:InfMemSI}}\label{Sec:InfMemSI}
As in the case without SI, when $Z_{t}^y=Y^t$, we can use the tools of MDP to derive a structure theorem. We will need to redefine the state to $s_t = P_{X_t,Z_{t-1}^w|Y^{t-1}}(\cdot,\cdot|y^{t-1})$. The action is defined in the same manner as in the case without SI, i.e., $a_t:\calX\to\calY$. The optimal reproduction function, $\hx^*_t=g^*(w_t,y^t,z_{t-1}^w)$ is the Bayes response to $P_{X_t|W_t,Y^t,Z_{t-1}^w}(\cdot|w_t,y^t,z_{t-1}^w)$:
\begin{align}
    \hx^*_{t} = \arg\max_{\hx}\sum_{x_t}P(x_t|w_t,y^t,z_{t-1}^w)\rho_t(x_t,\hx).
\end{align}

As in Subsection \ref{Sec:InfMemNoSI}, in order to use the tools of MDP, we need to show that we can write the cost function as a function of $(s_t,a_t)$ and that the state is conditionally Markov, given $a_t$.

The optimal reproduction function is a function of $P_{X_t|W_t,Y^t,Z_{t-1}^w}(\cdot|w_t,y^t,z_{t-1}^w)$. Note that
\begin{align}
    P(x_t|w_t,y^t,z_{t-1}^w)&=\frac{P(w_t,x_t,y_t,z_{t-1}^w|y^{t-1})}{\sum_{x'_t}P(w_t,x'_t,y_t,z_{t-1}^w|y^{t-1})}\nt\\
    &=\frac{P(x_t,z_{t-1}^w|y^{t-1})P(w_t|x_t)\ind{a_t(x_t)=y_t}}{\sum_{x'_t}P(x'_t,z_{t-1}^w|y^{t-1})P(w_t|x'_t)
    \ind{a_t(x'_t)=y_t}}\nt\\
    &\eqd f(s_{t},a_t,w_t,x_t,y_t,z_{t-1}^w)
\end{align}
Therefore, the optimal reproduction function is a function of $(s_{t},a_t,w_t,y_t,z_{t-1}^w)$, which we denote by $g_t^*(s_t,a_t,w_t,y_t,z_{t-1}^w)$.
We now move on to show that the cost function can be written as a function of the state and action.
As in Subsection \ref{Sec:InfMemNoSI}, we deal with the distortion and codeword length elements of the cost separately. Treating the expected codeword length first we have:
 \begin{align}
    L_{Y_t|Y^{t-1}}(y^{t-1}) &= \min_{l(\cdot)\in\calA}\sum_{y_t}P(y_t|y^{t-1})l(y_t)\nt\\
    &= \min_{l(\cdot)\in\calA}\sum_{x_t,y_t,z_{t-1}^w}P(x_t,y_t,z_{t-1}^w|y^{t-1})l(y_t)\nt\\
    &= \min_{l(\cdot)\in\calA}\sum_{y_t,x_t,z_{t-1}^w}P(x_t,z_{t-1}^w|y^{t-1})\ind{a_t(x_t)=y_t}l(y_t)\nt\\
    &\eqd \alpha_t(s_t,a_t),
\end{align}

When using the optimal reproduction function, the average distortion is given by:
\begin{align}
    &\bE\left[\rho(X_t,g^*_t(s_t,a_t,Y_t,W_t,Z_{t-1}^w))\bigg|Y^{t-1}=y^{t-1}\right]\nt\\
    &~~~~=\sum_{w_t,x_t,y_t,z_{t-1}^w}P(x_t,y_t,z_{t-1}^w|y^{t-1})\rho(x_t,g^*_t(s_t,a_t,w_t,y_t,z_{t-1}^w))\nt\\
    &~~~~=\sum_{w_t,x_t,y_t,z_{t-1}^w}P(x_t,z_{t-1}^w|y^{t-1})P(w_t|x_t)\ind{a_t(x_t)=y_t}\rho(x_t,g^*_t(s_t,a_t,w_t,y_t,z_{t-1}^w))\nt\\
    &~~~~\eqd\beta_t(s_t, a_t).
\end{align}
Denoting $\beta_t(s_t, a_t) +\lambda \alpha_t(s_t, a_t) = \gamma_t(s_t, a_t)$, our optimality criterion can be written as $\frT\sum_{t=1}^T\bE\gamma_t(s_t, a_t)$.

We move on to show that the state process is Markov conditioned on the action, namely, $P(s_{t+1}|s^{t},a^{t}) = P(s_{t+1}|s_{t},a_t)$.
We start by noting that $s_{t+1}=P_{X_{t+1},Z_{t}^w|Y^{t}}(\cdot,\cdot|y^{t})$ is a function of $(s_t,a_t,y_t)$. For every $(x_{t+1},z_{t}^w)$ we have
\begin{align}
    &P(x_{t+1},z_{t}^w|y^{t}) = \frac{\sum_{w_t,x_t,z_{t-1}^w}P(x_t,x_{t+1},w_t,y_t,z_{t-1}^w,z_{t}^w|y^{t-1})} {\sum_{w_t,x_t,x_{t+1},z_{t-1}^w}P(x_t,x_{t+1},w_t,y_t,z_{t-1}^w,z_{t}^w|y^{t-1})}\nt\\
    &= \frac{\sum_{w_t,x_t,z_{t-1}^w}P(x_t,z_{t-1}^w|y^{t-1})P(x_{t+1}|x_t)P(w_t|x_t)\ind{a_t(x_t)=y_t}\ind{z_t^w=r^w_t(w_t,y_t,z_{t-1}^w)}} {\sum_{w_t,x_t,x_{t+1},z_{t-1}^w}P(x_t,z_{t-1}^w|y^{t-1})P(x_{t+1}|x_t)P(w_t|x_t)\ind{a_t(x_t)=y_t}\ind{z_t^w=r^w_t(w_t,y_t,z_{t-1}^w)}}\nt\\
    &=f(s_t,a_t,x_{t+1},y_t,z_t^w)\label{eq:RecStateSI}
\end{align}
and therefore, $s_{t+1}=h(s_t,a_t,y_t)$ for a function, $h$, that uses \eqref{eq:RecStateSI} for every pair $(x_{t+1},z_{t}^w)$.
Now,
\begin{align}
    P(s_{t+1}=\nu|s^t,a^t)&=\sum_{y_t,x_t,z_{t-1}^w}\ind{h(a_t,s_t,y_t)=\nu}\ind{a_t(x_t)=y_t}P(x_t,z_{t-1}^w|y^{t-1})\nt\\
    &=P(s_{t+1}=\nu|s_t,a_t).
\end{align}
We showed that our system can be represented as a MDP. By invoking Theorem \ref{Thm:AppMDP}, we know that the optimal action at each stage, $a_t$ is a deterministic function of the state. Namely, the mapping from $x_t$ to $y_t$ can be chosen deterministically as a function of $P_{X_t,Z_{t-1}^w|Y^{t-1}}(\cdot,\cdot|y^{t-1})$. Therefore, $Y_t$ is a deterministic function of $(X_t,P_{X_t,Z_{t-1}^w|Y^{t-1}}(\cdot,\cdot|y^{t-1}))$, which concludes the proof of Theorem \ref{Thm:InfMemSI}.
By \eqref{eq:RecStateSI}, the encoder does not need to store $y^{t-1}$, but rather a probability measure (a vector in $\mathbbm{R}^{|\calY|\times|\calZ^w|}$).

\section{Conclusion} \label{Sec:Conclusion}
This work extended the setting of \cite{Witsenhausen1979} to include both variable rate coding and SI. It was shown that structure theorems, in the spirit of \cite{Witsenhausen1979} and \cite{Teneketzis2006}, continue to hold in this setting as well. These theorems are further refined when the decoder has infinite memory. We were able to show that the cost function is concave in the choices of past encoders (Lemmas \ref{Lem:2StageConcave}, \ref{Lem:3StageConcave}) and therefore, the optimal encoders are deterministic. It was also shown that in order to simplify the overall system optimization, one can use sliding--window next--state functions and the excess loss incurred by this suboptimal choice vanishes as the window size increases (Theorems \ref{Thm:MarkovMem}, \ref{Thm:Combined}). However, in the finite horizon setting we investigated, the window size is always upper bounded by the time horizon.

Extensions to this work would include investigating the infinite horizon setting. While Theorem \ref{Thm:MarkovMem} carries over verbatim to the infinite horizon setting, it is not necessarily true for the other theorems, which were proved using dynamic programming. Another extension would be to investigate the constrained setting (briefly mentioned in Theorem \ref{Thm:ConstrainedRate}). In this case, relying on results from constrained MDPs, we do not expect the optimal encoders to be deterministic (see \cite{AltmanBook99}).
\input{Appendix}

\bibliographystyle{IEEEtran}
\bibliography{PhDBib}

\end{document}

%% file: Defs.tex
\newcommand{\ba}{\begin{align*}}
\newcommand{\eaa}{\end{align*}}
\newcommand {\nt} {\notag}

\newcommand{\lb}{\left(}
\newcommand{\rb}{\right)}

\newcommand{\frT}{\frac 1 T}

\newcommand{\calX}{{\cal X}}
\newcommand{\calY}{{\cal Y}}
\newcommand{\calZ}{{\cal Z}}

\newcommand{\calA}{{\cal A}}
\newcommand{\calB}{{\cal B}}
\newcommand{\calW}{{\cal W}}

\newcommand{\calS}{{\cal S}}

\newcommand {\bE} {\mbox{\boldmath $E$}}

\newcommand {\hX}{\hat{X}}
\newcommand {\hx}{\hat{x}}

\newcommand {\hY}{\hat{Y}}
\newcommand {\hZ}{\hat{Z}}

\newcommand{\eqd}{\stackrel{\triangle}{=}}

\newcommand {\reals} {{\rm I\!R}}
\newcommand{\ind}[1] {\mathbbm{1}\left\{#1\right\}}
\newcommand{\sbr}[1] {\left[#1\right]}
\newcommand{\cbr}[1] {\left\{#1\right\}}
\newtheorem{theorem}{Theorem}
\newtheorem{lemma}{Lemma}
\newtheorem{corollary}{Corollary}

%% file: Introduction.tex
\section{Introduction}\label{Sec:Intro}
We consider the following source coding problem. Symbols produced by a discrete Markov source are to be encoded, transmitted noiselessly and reproduced by a decoder which can have causal access to side information (SI) correlated to the source. Operation is in real time, that is, the encoding of each symbol and its reproduction by the decoder must be performed without any delay and the distortion measure does not tolerate delays.

The decoder is assumed to be a finite--state machine with a fixed number of states. With no SI, the scenario where the encoder is of fixed rate was investigated by Witsenhausen \cite{Witsenhausen1979}. It was shown that for a given decoder, in order to minimize the distortion at each stage for a Markov source of order $k$, an optimal encoder can be found among those for which the encoding function depends on the $k$ last source symbols and the decoder's state (in contrast to the general case where its a function of all past source symbols).
Walrand and Varaiya \cite{WalrandVaraiya83} extended this finding to a joint
source--channel setup with noiseless feedback. Teneketzis \cite{Teneketzis2006} used ideas from both \cite{Witsenhausen1979} and \cite{WalrandVaraiya83} and considered the joint source--channel setup for a given finite state decoder without feedback. A causal variant of the Wyner Ziv problem \cite{WynerZiv76} was also considered by Teneketzis \cite{Teneketzis2006}. It is shown in \cite{Teneketzis2006} that the optimal (fixed rate) encoder for this case is a function of the current source symbol and the probability mass function of the decoder's state for the symbols sent so far. Borkar, Mitter and Tatikonda
\cite{BorkarMitterTatikonda2001} derived
structure theorems of a similar spirit when the cost function is a
linear combination (Lagrangian) of the conditional entropy of the
reproduction sequence and the mean square error of the reproduction.
The scenario where the encoder is also a finite state machine was considered by Gaarder and Slepian in \cite{GaarderSlepian1982}. In some cases, the minimization of the distortion (or cost) can be cast as a stochastic control problem. In this case, tools developed for Markov decision processes are employed to either solve the optimization problem or to get insights on the structure of the optimal solution. Examples of this technique include \cite{WalrandVaraiya83},\cite{Teneketzis2006},\cite{BorkarMitterTatikonda2001},\cite{GorantlaColeman2011},\cite{AsnaniWeissmann11}.

When the time horizon and alphabets are finite, there is a finite number of possible deterministic encoding, decoding and memory update rules. In principle, a brute force search would yield the optimal choice. However, since the number of possibilities increases doubly exponentially in the duration of the communication and exponentially in the alphabet size, it is not trackable even for very short time horizons. Recently, using the results of \cite{Teneketzis2006},  Mahajan and Teneketzis \cite{MahajanTeneketzis2009} proposed a search frame that is linear in the communication duration and doubly exponential in the alphabet size.

Real time codes are a subclass of causal codes, as defined by Neuhoff and Gilbert \cite{NeuhoffGilbert1982}. In \cite{NeuhoffGilbert1982}, entropy coding is used on the whole sequence of reproduction symbols, introducing arbitrarily long delays. In the real time case, entropy coding has to be instantaneous, symbol--by--symbol (possibly taking into account past transmitted symbols). It was shown in \cite{NeuhoffGilbert1982}, that for a discrete memoryless source (DMS), the optimal causal encoder consists of time--sharing between no more than two memoryless encoders. Weissman and Merhav \cite{TsachyNeri2005} extended \cite{NeuhoffGilbert1982} to the case where SI is also available at the decoder, encoder or both. Error exponents for real time coding with finite memory for a DMS where derived in \cite{NeriIoannis2003}.

This work extends \cite{Witsenhausen1979} in several directions: The first is extending the result of \cite{Witsenhausen1979} from fixed--rate coding to variable--rate coding, where accordingly, the cost function is redefined so as to incorporate both the expected distortion and the expected coding rate. Secondly, we allow the decoder access to causal side information. Unlike \cite{Witsenhausen1979} and \cite{Teneketzis2006}, we do not a-priori restrict the encoders to be deterministic and thus the encoders can be any stochastic function of all causally available data. While in \cite{Witsenhausen1979} and \cite{Teneketzis2006}, it is quite clear that deterministic encoders are a--priori optimal, it is not immediately clear in our case, as we discuss in the sequel. We show that structure theorems, in the same spirit as those of Witsenhausen \cite{Witsenhausen1979} and Teneketzis \cite{Teneketzis2006}, continue to hold in this setting as well. Moreover, the structure can be simplified when the decoder has infinite memory. Finally, we upper bound the loss incurred by using a suboptimal next--state function which uses a ``sliding--window'' over the past decoder inputs. We refer to such memory update functions as \textit{Markov memory update functions}. The upper bound is given in terms of the original state alphabet and the window length. The suboptimal system that uses Markov memory update functions is analytically more tractable and its optimization is easier since in order to find the best sub--optimal system, effectively, as discussed in the sequel, only the encoders need to be optimized.

In contrast to \cite{Witsenhausen1979} and \cite{Teneketzis2006}, where fixed--rate coding was considered, and hence the performance measure was just the expected distortion, here, since
we allow variable--rate coding, our cost function incorporates both rate
and distortion. This is done by defining our cost function in terms of
the Lagrangian
\begin{align*}
     \text{(distortion)} +\lambda\cdot\text{(code length)}.
\end{align*}
where $\lambda>0$ is a fixed parameter that controls the tradeoff between rate and distortion.
In \cite{Witsenhausen1979}, the proof of the structure theorem relied on two lemmas. The proofs of the extensions of those lemmas to our case are more involved than the proofs of their respective original versions in \cite{Witsenhausen1979}. To intuitively see why, remember that the proof of the lemmas in \cite{Witsenhausen1979}, relied on the fact that for every decoder state, source symbol and a given decoder, since there is a finite number of possible encoder outputs (governed by the fixed rate), we could choose the one minimizing the distortion. However, in our case, such a choice might entail a large expected coding rate, and although minimizes the distortion, it will not minimize the overall cost function (especially for large $\lambda$). Furthermore, unlike the case of \cite{Witsenhausen1979}, in our setting, the cost in future stages depends non--linearly on the choices of earlier encoders and in contrast to \cite{Witsenhausen1979} and \cite{Teneketzis2006}, there is no reason, as we discuss in the sequel, to a--priori assume that deterministic encoders are optimal.

The remainder of the paper is organized as follows: In Section \ref{Sec:Prelim}, we give the formal setting and notation used throughout the paper. In Section \ref{Sec:NoSI}, we start with the simpler setting without SI. Structure theorems regarding the encoder are derived for both the finite and infinite memory models. In Section \ref{Sec:MarkovMemUpdate}, we upper bound the loss incurred when  Markov memory functions are used instead of the optimal next--state functions. In Section \ref{Sec:SI}, we exend the setting of Section \ref{Sec:NoSI} by allowing the decoder access to SI. We begin each section by stating and discussing its main result. Finally, we conclude this work in Section \ref{Sec:Conclusion}. 

%% file: prelims.tex
\section{Preliminaries \label{Sec:Prelim}}
We begin with notation conventions. Capital letters represent scalar random variables (RV's), specific
realizations of them are denoted by the corresponding lower case letters, and their alphabet -- by calligraphic letters. For $i < j$ ($i$, $j$ –- positive integers), $x^j_i$ will denote the vector $(x_i,\ldots, x_j)$, where for $i = 1$ the subscript will be omitted. $P_X(\cdot)$ will denote a probability measure over $\calX$. When there is no room for ambiguity, we will use $P(x)$ instead of $P_X(x)$. $\ind{A}$ will denote the indicator of the event $A$.

We consider a Markov source producing a random sequence $X_1, X_2, ..., X_T$, $X_t\in \calX$, $t=1,2,\ldots,T$. The cardinality of $\calX$, as well as those of other alphabets in the sequel, is finite.
The probability mass function of $X_1$, $P(x_1)$ and the transition probabilities, denoted by $P(x_t|x_{t-1})$, $t=2,3,\ldots,T$ are known.

Let $\calY$ denote the index set $\{1,2,\ldots,M\}$ for some finite $M$.
A variable--length stochastic encoder is a sequence of functions $\left\{f_t\right\}_{t=1}^T$. At stage $t$, a stochastic encoder uses all the causally available data, $(X^t,Y^{t-1})$, to choose a probability measure over $\calY$ from which $Y_t$ is drawn. After drawing $Y_t$, the encoder noiselessly transmits an entropy--coded codeword of $Y_t$.
A deterministic encoder is a stochastic encoder which draws a specific $Y_t\in\calY$ with probability $1$ (i.e., $Y_t$ is a deterministic function of $(X^t,Y^{t-1})$).
Unlike the fixed rate regime in \cite{Witsenhausen1979},\cite{Teneketzis2006}, where $\log_2|\calY|$ (rounded up) was the rate of the code at stage $t$, here the subset of $\calY$ used at each stage, along with the length of the binary representation of $Y_t$, will be subject to optimization.

The encoder structure is not confined {\it a--priori}, and at each time instant $t$, $Y_t$ may be given by
an arbitrary (possibly stochastic) function of $(X^t,Y^{t-1})$ as described above. The decoder, however,
is assumed, similarly as in \cite{Witsenhausen1979} and \cite{Teneketzis2006}, to be a finite--memory device,
defined as follows: At each stage, $t$, the decoder updates its current state (or memory) and outputs a reproduction symbol $\hX_t$.
We assume that the decoder state, $Z_t$, is updated by
\begin{align}
    Z_1 &= r_1(Y_1)\nt\\
    Z_t &= r_t(Y_t, Z_{t-1}), ~~~t=2,3,\ldots,T
\end{align}
Since the transmission is noiseless, $Z_t$ can be tracked by the encoder. Note that this model also includes infinite memory, i.e., $Z_t=Y^t$. The reproduction symbols are produced by a sequence of functions $\left\{g_t\right\}$, $g_t:\calY\times\calZ\to\hat{\calX}$ as follows
\begin{align}
    \hX_1 &= g_1(Y_1)\nt\\
    \hX_t &= g_t(Y_t, Z_{t-1}), ~~~t=2,3,\ldots,T
\end{align}

Since at the beginning of stage $t$, $Z_{t-1}$ is known to both encoder and decoder, the entropy coder at every stage needs to encode the random variable $Y_t$ given $Z_{t-1}=z_{t-1}$. We define $\calA$ to be the set of all instantaneously uniquely decodable codes for $\calY$, i.e., all possible length functions $l:\calY\to \mathbb{Z}_+\cup\infty$ that satisfy Kraft's inequality:
\begin{align}
    \calA = \left\{l(\cdot):\sum_{y\in\calY}2^{-l(y)}\leq 1 \right\}.
\end{align}
Note that we allow infinite--length codewords. We will return to this technical issue after properly defining the cost function.
The average codeword length at stage $t$, for a specific decoder state $z_{t-1}$, will be given by:
\begin{align}
    L_{Y_t|Z_{t-1}}(z_{t-1}) \eqd \left\{ \begin{array}{ll}
    0 & \text{if }\max_{y_t\in\calY}P(y_t|z_{t-1})=1\\
    \min_{l(\cdot)\in\calA}\left\{\sum_{y_t\in\calY}P(y_t|z_{t-1})l(y) \right\} & \text{otherwise}
    \end{array}\right.\label{eq:LYGivenZ}.
\end{align}
i.e., if given $Z_{t-1}=z_{t-1}$, $Y_t$ is deterministically known, there is no need to transmit any information, otherwise
$L_{Y_t|Z_{t-1}}(z_{t-1})$ is obtained by designing a Huffman code for the probability distribution $P_{Y_t|Z_{t-1}}(\cdot|z_{t-1})$.
Note that for given encoders and state update functions, $L_{Y_t|Z_{t-1}}(z_{t-1})$ is a function of $z_{t-1}$ only.
Also, $L_{Y_t|Z_{t-1}}(z_{t-1})$ is discontinuous around $0$ in the distribution $P_{Y_t|Z_{t-1}}(\cdot|z_{t-1})$ since if given $Z_{t-1}=z_{t-1}$, $Y_t$ is not deterministically known, then $L_{Y_t|Z_{t-1}}(z_{t-1})\geq 1$.

\begin{figure}[htp]
\centering
\includegraphics[width=0.8\textwidth]{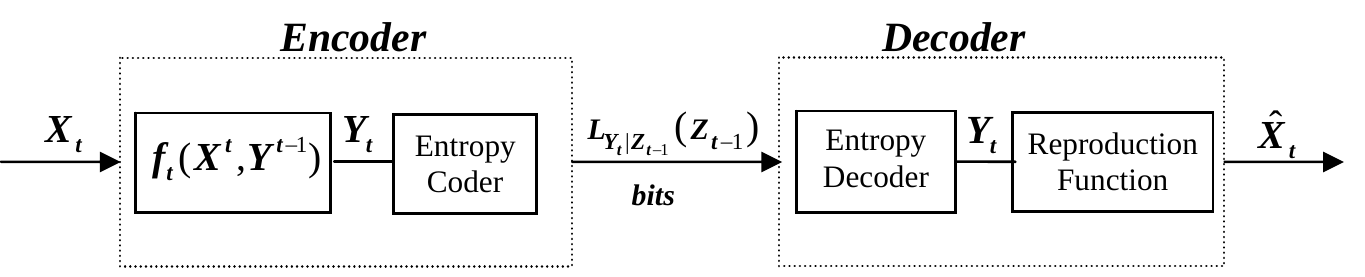}
\caption{System model \label{Fig:Model}}
\end{figure}

The average codeword length of stage $t$, denoted $L_{Y_t|Z_{t-1}}$, is defined as $\bE L_{Y_t|Z_{t-1}}(Z_{t-1})$, where the expectation is with respect to $Z_{t-1}$.
Our system model is depicted in Figure \ref{Fig:Model}.

We are given a sequence of distortion measures $\left\{\rho_t\right\}_{t=1}^T$, $\rho_t:\calX\times\hat{\calX}\to \reals^{+}$.  At each stage, the cost function is a linear combination of the average distortion and codeword length $L_{Y_t|Z_{t-1}}$, i.e.,
\begin{align}
    J_t \eqd \bE\left\{\rho_t(X_t, \hX_t) + \lambda L_{Y_t|Z_{t-1}}(Z_{t-1})\right\}\label{eq:J_t},
\end{align}
where $\lambda>0$ is a fixed parameter that controls the tradeoff between rate and distortion.
Our goal is to minimize the average cost
\begin{align}
J \eqd \frac 1 T\sum_{t=1}^T J_t.\label{eq:TotalCost}
\end{align}

A sequence of encoders, $f_1,\ldots,f_T$, is said to be optimal if for a given sequence of decoders and memory update functions, $f_1,\ldots,f_T$ attains $\inf J$, where the infimum is over the set of all sequences of stochastic encoders, which are functions of all causally available data.

A stage--$t$ encoder is said to be optimal if given the future stages encoders and decoders, it attains $\inf\sum_{i=t}^T J_i$, where the infimum is over the set of stochastic stage--$t$ encoders (which are functions of $(X^t,Y^{t-1})$).

Note that for large enough $\lambda$, for some $z_{t-1}$, the optimal encoders might use only a small subset of $\calY$ (thus attaining higher distortion but smaller overall cost). Technically, this means that there will be a subset $\calB\subset\calY$ such that $P(y_t|z_{t-1})=0$ if $y_t\in\calB^c$. We therefore need that $\calA$ will contain good codes for subsets of $\calY$. By allowing infinite length codewords, we make sure that $\calA$ contains codes which are uniquely decodable for all subsets of $\calY$ (and satisfy Kraft's inequality for alphabet $\calY$). Needless to say that with this definition, a code for $\calB\subset\calY$ will be used iff $P(y_t|z_{t-1})=0$ for all $y_t\in\calB^c$, where we use $0\cdot\infty=0$.

%A specific choice of the memory update functions $\left\{r_t\right\}$, reproduction functions $\left\{g_t\right\}$ and encoders $\left\{f_t\right\}$ along with their resulting cost $J$ is called a \textit{design}. We say that design $A$, with cumulative cost $J^A$, outperforms design $B$, with cumulative cost $J^B$, if $J^A\leq J^B$.

%% file: NoSI.tex
\section{Structure Theorems - No Side Information}\label{Sec:NoSI}
\subsection{Main results}
We start by briefly stating and discussing the main contributions of this section. The proofs of the following theorems are found in the following subsections.

The first contribution of this paper is the following theorem, which basically states that the results of \cite{Witsenhausen1979} continue to hold in this setting as well.
\begin{theorem}\label{Thm:NoSI}
    For a Markov source and any given sequence of memory update functions $\{r_t\}$, reproduction functions $\{g_t\}$ and distortion measures $\{\rho_t\}$, there exists a sequence of deterministic encoders $Y_t = f_t(X_t,Z_{t-1})$ which is optimal.
\end{theorem}
The addition of the variable--rate coding and allowing a larger class of encoders compared to \cite{Witsenhausen1979}, makes the proof of this result considerably more involved than its counterpart in \cite{Witsenhausen1979}, as was discussed at the end of Section \ref{Sec:Intro}.

While Theorem \ref{Thm:NoSI} covers the infinite decoder memory ($Z_t=Y^t$) setting, in this case, when optimal reproduction functions are used (see Section \ref{Sec:InfMemNoSI}), we have the following theorem, which refine Theorem \ref{Thm:NoSI} for this case:
\begin{theorem}\label{Thm:InfMem}
    For a Markov source and any sequence of distortion measures $\{\rho_t\}$ and optimal infinite memory decoders, there exists a sequence of deterministic encoders $Y_t=f_t(X_t,P_{X_t|Y^{t-1}}(\cdot|y^{t-1}))$ which is optimal.
\end{theorem}
We will show that $P_{X_t|Y^{t-1}}(\cdot|y^{t-1})$ can be recursively updated. Theorem \ref{Thm:InfMem} is a refinement of Theorem \ref{Thm:NoSI} since, in the setup of Theorem \ref{Thm:InfMem}, there is no need to store the whole history of encoder outputs, $Y^t$, as the statement of Theorem \ref{Thm:NoSI}, but instead, $P_{X_t|Y^{t-1}}(\cdot|y^{t-1})$ is recursively updated. (given that a probability measure can be stored).

In the remainder of this section, we will prove Theorems \ref{Thm:NoSI} and \ref{Thm:InfMem}, starting with Theorem \ref{Thm:NoSI}.
In order to prove Theorem \ref{Thm:NoSI}, we need a few supporting lemmas, as in \cite{Witsenhausen1979}. In the following two subsections, we state and prove the supporting lemmas and then prove Theorem \ref{Thm:NoSI} in Subsection \ref{Sec:ProofThm1}. Theorem \ref{Thm:InfMem} is proved in Subsection \ref{Sec:InfMemNoSI}.

%The proof of this theorem \ref{Thm:NoSI}, as in \cite{Witsenhausen1979}, requires a few auxiliary lemmas. While the lemmas in \cite{Witsenhausen1979}, make similar assumptions as our lemmas, their proofs in our case follows a different track. In \cite{Witsenhausen1979}, Witsenhausen's principal tool in the proof is the fact that since the alphabet of $Y$ is finite, for every pair $(x_t,z_{t-1})$, there is a specific $y_t$ that minimizes $\rho_t(x_t,g_t(y_t,z_{t-1}))$ (or some other function of $x_y,y_t,z_{t-1}$). The optimal encoder chooses this $y_t$ for every pair $(x_t,z_{t-1})$. This method is applied before taking the expectation on the distortion measure. In our case however, this approach can not be taken since the length function, depends on the distribution $P(y_t|z_{t-1})$, and therefore a symbol--by--symbol approach can no be taken here. Intuitively, while Witsenhausen's choice of encoder minimizes the distortion element of the cost, it is not clear how such a choice will influence $P(y_t|z_{t-1})$ and  it might entail a large average length. Moreover, while in \cite{Witsenhausen1979} it is clear that the encoders should be deterministic, it is not immediately clear that this is true in out case as well. We further discuss this issue in the following.
%
%In the following three subsections, we state and prove the auxiliary lemmas. Using these lemmas, we prove theorem \ref{Thm:NoSI} in subsection \ref{Sec:ProofThm1}.

\subsection{Two--stage lemma}\label{Sec:2StageLemma}
We start by analyzing a system with two stages only, where the first encoder is known. 
\begin{lemma}\label{Lem:2Stage}
    For any two--stage system ($T=2$), there exists a deterministic second stage encoder $Y_2=f_2(X_2,Z_1)$, which is optimal.
\end{lemma}
\noindent {\it Proof:} Note that $f_1,g_1,g_2,r_1$ are fixed, and so, $J_1$ is unchanged by changing $f_2$. We need to show that a second stage encoder, that minimizes $J_2$, can be a deterministic function of $(X_2, Z_1)$. Denote the set of stochastic encoders which are functions of $(X_1,X_2,Y_1)$ by $\{f_{X^2Y_1}^s \}$.
For every joint probability measure over the quadruple $(X_1,X_2,Y_2,Z_1)$, $J_2$ is well defined and our objective is to find the optimal encoder that attains:
\begin{align}
    \inf_{\{f_{X^2Y_1}^s \}}J_2 = \inf_{\{f_{X^2Y_1}^s \}} \bE \left\{\rho_2(X_2, g_2(Y_2,Z_1))+ \lambda L_{Y_2|Z_1}(Z_{1})\right\} \label{eq:Minimization}.
\end{align}
Consider the random quintuple $(X_1,X_2,Y_1,Y_2,Z_1)$ which takes part in the expectation of \eqref{eq:Minimization}. From the structure of the system, we know that
\begin{align}
    P(x_1,x_2,y_1,y_2,z_1)=P(x_1)P(x_2|x_1)P(y_1|x_1)P(y_2|x_1,x_2,y_1)\ind{r_1(y_1)=z_1}\label{eq:5Tuple},
\end{align}
where we used the fact that $z_1$ is a deterministic function of $y_1$. Everything but the second stage encoder, which directly affects $P(y_2|x_1,x_2,y_1)$ is fixed. Note that the optimization affects $L_{Y_2|Z_1}(Z_{1})$ since
\begin{align}
    L_{Y_2|Z_1}(z_1) = \min_{l(\cdot)\in\calA}\sum_{y_2}\sum_{x_2}P(x_2|z_1)P(y_2|x_2,z_1)l(y_2)\label{eq:2StgTmp}
\end{align}
and $P(y_2|x_2,z_1)$ depends on $P(y_2|x_1,x_2,y_1)$ as we will show shortly.

Let $\{f_{X_2Z_1}^s \}$ denote the subset of stochastic encoders which are functions of $(X_2,Z_1)$. Also, let $\{f_{X_2Z_1}^d \}$ denote the subset of \textit{deterministic} encoders which are functions of $(X_2,Z_1)$. Since $Z_1$ is a function of $Y_1$, $\{f_{X_2Z_1}^d \} \subset \{f_{X_2Z_1}^s \}\subset\{f_{X^2Y_1}^s \}$. We prove Lemma \ref{Lem:2Stage} in two steps. First, we show that it is enough to search in the (infinite) subset $\{f_{X_2Z_1}^s \}$. In the second step, we show that among $\{f_{X_2Z_1}^s \}$, the optimal encoder is a member of $\{f_{X_2Z_1}^d \}$.

\textit{Step 1:} We rewrite \eqref{eq:Minimization} as follows:
\begin{align}
    \inf_{\{f_{X^2Y_1}^s \}} J_2 &= \inf_{\{f_{X^2Y_1}^s \}} \sum_{x_2,y_2,z_1} P(x_2,y_2,z_1)\sbr{\rho_2(x_2, g_2(y_2,z_1))+ \lambda L_{Y_2|Z_1}(z_{1})}\nt\\
    &= \inf_{\{f_{X^2Y_1}^s \}} \sum_{x_2,y_2,z_1} P(x_2,y_2,z_1)\times\nt\\
    &\sbr{\rho_2(x_2, g_2(y_2,z_1))+ \lambda \min_{l(\cdot)\in\calA_t}\sum_{y'_2}\sum_{x'_2}P(x'_2|z_1)P(y'_2|x'_2,z_1)l(y'_2)}\label{eq:ENoSI}.
\end{align}
Now, given that the first stage encoder and decoder are known, $P(x_2,z_1)$ is well defined since
\begin{align}
    P(x_2,z_1) &= \sum_{x_1,y_1}P(x_1,x_2,y_1,z_1)\nt\\
    &= \sum_{x_1,y_1}P(x_1)P(x_2|x_1)P(y_1|x_1)\ind{r_1(y_1)=z_1}
\end{align}
and $P(x_1),P(x_2|x_1),\ind{r_1(y_1)=z_1}$ are determined by the known source and first stage next--state function, $P(y_1|x_1)$ is directly determined by the first stage encoder.
Also, by the Bayes rule, we have, for any second stage encoder:
\begin{align}
    P(x_2,y_2,z_1)=P(y_2|x_2,z_1)P(x_2,z_1). \label{eq:PY2Z1X2}
\end{align}
Therefore,
\begin{align}
    \inf_{\{f_{X^2Y_1}^s \}} J_2 &= \inf_{\{f_{X^2Y_1}^s \}} \sum_{x_2,y_2,z_1} P(y_2|x_2,z_1)P(x_2,z_1)\times\nt\\
    &\sbr{\rho_2(x_2, g_2(y_2,z_1))+ \lambda \min_{l(\cdot)\in\calA_t}\sum_{y'_2}\sum_{x'_2}P(x'_2|z_1)P(y'_2|x'_2,z_1)l(y'_2)}\label{eq:2StgStart}.
\end{align}
The only term that is affected by the optimization is $P(y_2|x_2,z_1)$. Observe that by \eqref{eq:5Tuple}, we have
\begin{align}
    P(y_2|x_2,z_1)&=\sum_{x_1,y_1}\frac {P(x_1,x_2,y_1,y_2,z_1)} {P(x_2,z_1)}\nt\\
    &= \sum_{x_1,y_1}\frac {P(x_1)P(x_2|x_1)P(y_1|x_1)P(y_2|x_1,x_2,y_1)\ind{r_1(y_1)=z_1}} {P(x_2,z_1)} \label{eq:beyes}.
\end{align}
From \eqref{eq:2StgStart}, \eqref{eq:beyes}, it is evident that the role of the second stage encoder in a two stage system is to select $P_{y_2|x_2,z_1}(\cdot|x_2,z_1)$ for every $(x_2,z_1)$ so as to minimize the cost. To see this, note that every $f_2\in\{f_{X^2Y_1}^s \}$ is mapped by \eqref{eq:beyes} (through $P(y_2|x_1,x_2,y_1)$ for every $(x_1,x_2,y_1)$) to a point on the simplex of probability measures on $\calY$ for every $(x_2,z_1)$. Namely, every $f_2\in\{f_{X^2Y_1}^s \}$ is mapped to $\hat{f}_2\in\{f_{X_2Z_1}^s \}$ and the optimization is affected only by $\hat{f}_2$. If instead of using a specific $f_2$ we will use $\hat{f}_2$ that results from it through \eqref{eq:beyes}, the joint probability $P(x_2,y_2,z_1)$ will remain the same and therefore, also the second stage cost.
Also note that we cannot gain anything from optimizing only over $\{f_{X_2Z_1}^s \}$ and not $\{f_{X^2Y_1}^s \}$ since $\{f_{X_2Z_1}^s \}$ is completely covered by $\{f_{X^2Y_1}^s \}$ through \eqref{eq:beyes}.
Therefore, since the optimization over $\{f_{X^2Y_1}^s \}$ is mapped to an optimization over $\{f_{X_2Z_1}^s \}$, we have
\begin{align}
    \inf_{\{f_{X^2Y_1}^s \}} J_2 = \inf_{\{f_{X_2Z_1}^s \}} J_2 \label{eq:AlternativeMin}
\end{align}
which completes the fist step of the proof.\\
\textit{Step 2:}
To complete the proof of the two--stage lemma, we need to show that it is enough to search in the finite space of \textit{deterministic} encoders which are functions of $(X_2,Z_1)$. Observe that the set of stochastic encoders is a convex set. The extreme points of this set (the points that are not convex combinations of other points) are deterministic encoders, namely, the set $\{f_{X_2Z_1}^d\}$. To complete the proof, we use the following lemma, proved in Appendix \ref{App:2ndCostConcave}.
\begin{lemma}\label{Lem:2StageConcave}
    The stage $t$ loss function is concave in $\{f_{X_tZ_{t-1}}^s\}$.
\end{lemma}
Using Lemma \ref{Lem:2StageConcave}, we conclude that since we minimize a concave function over a convex set, the minimizer will be one of the extreme points of the set, i.e., a member of $\{f_{X_2Z_1}^d \}$. We thus showed that
\begin{align}
    \inf_{\{f_{X^2Y_1}^s \}} J_2 = \min_{\{f_{X_2Z_1}^d \}} J_2
\end{align}

This completes the proof of the two--stage lemma. Note that no assumptions on the statistics of the source were made in the proof and therefore, the two--stage lemma holds for any source.\IEEEQED

\textit{Discussion:} \\
\textbf{1.} Observe that the actual optimal encoding function for each $(x_2,z_1)$ depends on the encoder of the first stage through $P(x_2,z_1)$ (which also governs $P(x_2|z_1)$), as seen from \eqref{eq:2StgStart}. This is true in general and not only in a two--stage system. The joint distribution $P_{X_t,Z_{t-1}}(\cdot,\cdot)$ can be thought of as the state of the system, governed by the choices of previous encoders (note however, that this state is static in the sense that it is not influenced by the actual realization of the source sequence). Therefore, the role of the stage $t$ encoder, besides greedily minimizing the stage $t$ cost (given the state $P_{X_t,Z_{t-1}}(\cdot,\cdot)$), is to control the future states so that they will allow minimal costs in future stages. This is true for all but the last encoder, which does not affect future cost, as seen for the second stage encoder in a two stage system. We will come back to this issue in Subsection \ref{Sec:InfMemNoSI} when we deal with infinite memory decoders and apply tools of stochastic control.\\
\textbf{2.} It is not surprising that the optimal second stage cost is attained by a deterministic encoder. Since the second stage is the last stage, the last encoder does not affect future costs and therefore, instead of using a convex combination of deterministic encoders (i.e., a stochastic encoder), use only the one with the best performance. However, in a system with more stages, it is not immediately clear that deterministic encoders in intermediate stages are optimal.
In fact, this is also true for the first stage of a two stage system. We saw that the first stage affects the second stage cost through $P(x_2,z_1)$. Specifically, it affects the second stage cost through $P(x_2|z_1)$, (as seen in \eqref{eq:ENoSI}) which is non linear in the first stage encoder $P(y_1|x_1)$ since
\begin{align}
    &P(x_2|z_1)=\frac{\sum_{x_1,y_1}P(x_1,x_2)P(y_1|x_1)\ind{r_1(y_1)=z_1}} {\sum_{x_1,x_2,y_1}P(x_1,x_2)P(y_1|x_1)\ind{r_1(y_1)=z_1}}.
\end{align}
If the first--stage encoder is deterministic, there is only a finite number of possible $P_{X_2|Z_1}(\cdot,\cdot)$.
Assume that we use a stochastic first--stage encoder, $f_1^s$. Although by Lemma \ref{Lem:2StageConcave}, $f_1^s$ is sub--optimal for $J_1$, can it allow us to reach a $P'_{X_2|Z_1}(\cdot,\cdot)$, unreachable by deterministic encoders, that will be favorable in terms of $J_2$ and yield a lower overall cost? We show in the sequel that the answer is negative and that the optimal first--stage encoder is deterministic as well. We will show that the \textit{stage $t$ cost is a concave functional in the choices of the previous stages encoders}. The proof of the last statement is much more involved than the proof of Lemma \ref{Lem:2StageConcave} and it is discussed in the next subsection. In \cite{Witsenhausen1979}, \cite{Teneketzis2006}, the stage $t$ distortion is linear in the choice of the encoders at all previous stages (since the expectation is linear and the non--linear element of the codeword length was not present). Therefore, there was no loss of optimality in a-priori confining the encoders to be deterministic. We further address this issue in the following subsection which deal with a more complex system.

\begin{corollary}\label{Cor:LastStage}
    In any T--stage system ($T\geq 2$) there exists a deterministic last stage encoder $Y_T=f_T(X_T, Z_{T-1})$, which is optimal.
\end{corollary}
\noindent {\it Proof:} Let $\hX_1 \eqd (X_1,X_2,\ldots,X_{T-1}),\hX_2\eqd X_T, \hZ_1=Z_{T-1}$, where $\hZ_1$ is calculated recursively according the the encoding functions that operate on $\hX_1$ and the resulting $Y_1,\ldots,Y_{T-1}$. We now apply the two--stage lemma to this system to conclude that the last stage encoder is a deterministic function of $(X_t, Z_{T-1})$.
\IEEEQED
\subsection{Three--stage lemma}
\begin{lemma}\label{Lem:3Stage}
    In a three-stage system $(T=3)$ with a Markov source, if the third--stage encoder is a deterministic function of $(X_3,Z_2)$, then there exists a deterministic second stage encoder $Y_2=f_2(X_2,Z_1)$, which is optimal.
\end{lemma}
\textit{Proof of Lemma \ref{Lem:3Stage}: }
We define, as in Subsection \ref{Sec:2StageLemma}, $\{f_{X^2Y_1}^s\}$ to be the set of all possible stochastic second--stage encoders. Let $\{f_{X_2Z_1}^s\}\subset\{f_{X^2Y_1}^s\}$ be the set that contains all stochastic second stage encoders that are functions of $(X_2,Z_1)$ and finally, let $\{f_{X_2Z_1}^d\}\subset\{f_{X_2Z_1}^s\}$ denote the set of deterministic encoders which are functions of $(X_2,Z_1)$.
Since the first--stage is fixed, $J_1$ is unaffected by changing the second stage encoder. Our goal is to jointly optimize $(J_2+J_3)$ with respect to the second stage encoder and show that
\begin{align}
    \inf_{\{f_{X^2Y_1}^s\}}\lb J_2+J_3\rb = \min_{\{f_{X_2Z_1}^d\}}\lb J_2+J_3 \rb.\label{eq:3StageGoalNoSI}
\end{align}
Since the third stage encoder is known, the expected third stage cost for any second stage encoder is given by
\begin{align}
    J_3 &= \bE\left\{\rho(X_3, g(Y_3,Z_2))+L_{Y_3|Z_2}(Z_2) \right\}\nt\\
    &= \sum_{x_3,y_3,z_2}P(x_3,z_2)P(y_3|x_3,z_2)\sbr{  \rho(x_3, g(y_3,z_2))+L_{Y_3|Z_2}(z_2)}\nt\\
    &= \sum _{x_3,y_3,z_2}P(x_3,z_2)\ind{f_3(x_3,z_2)=y_3} \rho(x_3, g(y_3,z_2)) + \sum_{z_2}P(z_2)\min_{l(\cdot)\in\calA}\sum_{x_3,y_3}\ind{f_3(x_3,z_2)=y_3}P(x_3|z_2)l(y_3).\label{eq:D3}
\end{align}
The second--stage encoder affects the last expression through $P(x_3,z_2)$ (and thus also through $P(z_2)$ and $P(x_3|z_2)$) since
\begin{align}
    P(x_3,z_2) &= \sum_{x_2,y_2,z_1}P(x_2,x_3,y_2,z_1,z_1)\nt\\
    &= \sum_{x_2,y_2,z_1} P(x_2,z_1)P(y_2|x_2,z_1)P(x_3|x_2)\ind{r_2(y_2,z_1)=z_2}\label{eq:PX3Z2}
\end{align}
where $P(x_2,z_1)$ is the result of the first--stage and we used the fact that the source is Markov and that $z_2$ is a deterministic function of $(y_2,z_1)$. Therefore, as we saw in Subsection \ref{Sec:2StageLemma}, the optimization affects the third--stage only through $P(y_2|x_2,z_1)$ for all $(x_2,y_2,z_1)$. We saw in \eqref{eq:2StgStart} that the second stage cost can be written as:
\begin{align}
    J_2 &= \sum_{x_2,y_2,z_1} P(y_2|x_2,z_1)P(x_2,z_1)\times\nt\\
    &\sbr{\rho_2(x_2, g_2(y_2,z_1))+ \lambda \min_{l(\cdot)\in\calA_t}\sum_{y'_2}\sum_{x'_2}P(x'_2|z_1)P(y'_2|x'_2,z_1)l(y'_2)}.
\end{align}
where $P(x_2,z_1)$ and thus $P(x_2|z_1)$ are the result of the first--stage encoder. We see that the optimization in the l.h.s of \eqref{eq:3StageGoalNoSI} affects both the second and third stage costs only through the conditional probabilities $P(y_2|x_2,z_1)$, for all $(x_2,y_2,z_1)$. Repeating the arguments used in the proof of Lemma \ref{Lem:2Stage}, instead of using a specific $f_2\in\{f_{X^2Y_1}^s\}$, we can use $\hat{f}_2\in\{f_{X_2Z_1}^s\}$ that results from it through \eqref{eq:beyes}, to draw $Y_2$. Since $P(x_2,y_2,z_1)$ will remain the same,
\begin{align}
P(x_3,z_2) = \sum_{x_2,y_2,z_1}P(x_2,z_1)P(y_2|x_2,z_1)P(x_3|x_2)\ind{r_2(y_2,z_1)=z_2}
\end{align}
will also remain the same and $\lb J_2+J_3 \rb$ will not be affected by this step.
We therefore have
\begin{align}
    \inf_{\{f_{X^2Y_1}^s\}} \lb J_2+J_3\rb =\inf_{\{f_{X_2Z_1}^s\}} \lb J_2+J_3 \rb. \label{eq:AlmostDone}
\end{align}
As in the two stage lemma, we need to show that it is enough to search in the space of deterministic encoders, $\{f_{X_2Z_1}^d\}$. Here, we have to show that both the second stage cost \textit{and} the third stage cost are concave in $\{f_{X_2Z_1}^s\}$. We know that the second stage cost is concave in $\{f_{X_2Z_1}^s\}$ from Lemma \ref{Lem:2StageConcave}. The following lemma asserts that the third stage cost is concave in $\{f_{X_2Z_1}^s\}$.
\begin{lemma}\label{Lem:3StageConcave}
    The third stage cost, $J_3$, is concave functional of $\{f_{X_2Z_1}^s\}$.
\end{lemma}
The proof Lemma \ref{Lem:3StageConcave} is much more involved than the proof of Lemma \ref{Lem:2StageConcave} and can be found in Appendix \ref{App:3StageConcave}.

Using lemma \ref{Lem:3StageConcave}, we conclude that $(J_2+J_3)$, which is the sum of two concave functionals, is concave in $\{f_{X_2Z_1}^s\}$. Therefore, the minimizer will be one of the extreme points of the convex set of $\{f_{X_2Z_1}^s\}$, namely, a member of $\{f_{X_2Z_1}^d\}$. We showed that
\begin{align}
    &\inf_{\{f_{X_2Z_1}^s\}} \lb J_2+J_3\rb = \min_{\{f_{X_2Z_1}^d\}} \lb J_2+J_3 \rb \label{eq:AltD3}
\end{align}
Using \eqref{eq:AlmostDone}, we arrive at \eqref{eq:3StageGoalNoSI} which completes the proof of Lemma \ref{Lem:3Stage}.
\IEEEQED

\subsection{Proof of Theorem \ref{Thm:NoSI} \label{Sec:ProofThm1}}
With the two-- and three--stage lemmas, we can prove Theorem \ref{Thm:NoSI} by using the method of \cite{Witsenhausen1979}, used for fixed rate encoding.
Theorem \ref{Thm:NoSI} is proven by backward induction. First apply Corollary \ref{Cor:LastStage} to any system to conclude that the optimal $f_T$ is a deterministic function of $(X_T,Z_{T-1})$. Now assume that the last $m$ encoders $f_{T-m+1},...,f_{T}$ are deterministic functions of $(X_{T-m+1},Z_{T-m}),\ldots,(X_{T},Z_{T-1})$, respectively. We will show that the encoder at time $(T-m)$ also has this structure and continue backwards until $t=2$. The first encoder is trivially a function of $X_1$ and by lemma \ref{Lem:3StageConcave} (with $Z_0$ as a constant) it is also deterministic. Let
\begin{align}
    \hX_1 &= (X_1,X_2,..., X_{T-m-1}),\nt\\
    \hY_1 &= (Y_1,Y_2,..., Y_{T-m-1}),\nt\\
    \hZ_1 &= \hat{r}_1(\hY_1),\nt\\
    \hX_2 &= X_{T-m},\nt\\
    \hY_2 &= Y_{T-m},\nt\\
    \hZ_2 &= r_{T-m}(\hY_{2},\hZ_1),\nt\\
    \hX_3 &= (X_{T-m+1}, X_{T-m+2},...,X_T),\nt\\
    \hY_3 &= (Y_{T-m+1}, Y_{T-m+2},...,Y_T),\label{eq:induction}
\end{align}
where $\hZ_1$ is recursively calculated from $\hY_1$ and it represents the state of the decoder after $(T-m-1)$ stages.
Using this new notation, the encoder that produces $\hY_3$ is a deterministic function of $(\hX_3,\hZ_2)$ (since, by assumption, the last $m$ encoders have the desired structure). The source is Markov since $\hX_3$ is independent of $\hX_1$ given $\hX_2$ (since the original source is Markov). Now, by the three--stage lemma, $\hY_2=Y_{T-m}=f_{T-m}(\hX_2,\hZ_1^y)=f_{T-m}(X_{T-m},Z_{T-m-1})$. Thus, the induction step is proved. This completes the proof of Theorem \ref{Thm:NoSI}.\IEEEQED

\textit{Remark:} Theorem \ref{Thm:NoSI} can be extended to a $k$-order Markov source using Witsenhausen's method \cite{Witsenhausen1979}. Namely, for a $k$-order Markov source, define $\tilde{X_1}=(X_1,X_2,\ldots,X_k)$, $\tilde{X_2}=(X_2,X_3,\ldots,X_{k+1})$ and so on. Now, $\tilde{X}_t$ is a Markov source. Using Theorem \ref{Thm:NoSI}, we can conclude that the optimal encoder is a function of the last $k$ source symbols and the state of the decoder.

\subsection{Infinite memory decoder - proof of Theorem \ref{Thm:InfMem}}\label{Sec:InfMemNoSI}
In this section, we deal with the case where the decoder has infinite memory, i.e., $Z_t = Y^{t}$. The memory update functions $\{r_t\}$ in this case are only appending the new received index $Y_t$ to $Z_{t-1}$. Note that this scenario is covered by Theorem \ref{Thm:NoSI}, however, in this case we can be more specific regarding the role of $Y^t$ at the encoder. While Theorem \ref{Thm:NoSI} was true for any decoding rule, Theorem \ref{Thm:InfMem} is true only for the optimal reproduction function. We define the \textit{Bayes Envelope} as
\begin{align}
B(P_{X_t|Y^t})\eqd\min_{\hat{x}_t}\sum_{x_t}P(x_t|y^{t})\rho_t(x_t,\hat{x}_t).
\end{align}
The minimizer of the last expression is called the \textit{Bayes--response} and will be denoted by $\hat{X}_{Bayes}(P_{X_t|Y^t})$. Clearly, $\hat{X}_{Bayes}(P_{X_t|Y^t})$ is a function of $P_{X_t|Y^t}(\cdot|y^t)$ and the cost function, $\rho_t$. The fact that the optimal reproduction function is the Bayes--response was shown in many places, for example \cite{Teneketzis2006},\cite[Lemma 3]{AsnaniWeissmann11}.

When infinite memory is available, we can use tools from Markov decision processes (MDP's) in order to derive a structure theorem. In Appendix \ref{App:MDPs}, we provide a brief background on MDP's. By Theorem \ref{Thm:NoSI}, we know that we can confine the discussion to deterministic encoders without loss of optimality. We need to show that our original problem can be represented as a MDP. The proof of Theorem \ref{Thm:InfMem} will follow immediately from Theorem \ref{Thm:AppMDP}, given in Appendix \ref{App:MDPs}. In order to show that we have an MDP, as we discuss in Appendix \ref{App:MDPs}, we need to show that:
\begin{itemize}
\item We can a find a sequence of deterministic functions $\{\gamma_t\}$, along with two finite spaces, $\calS,\calA$, such that the average cost, defined by \eqref{eq:J_t},\eqref{eq:TotalCost}, can be written as $J=\frT\bE\sum_{t=1}^T\gamma_t(s_t,a_t)$, where $s_t\in\calS, a_t\in\calA$ are the system state and the action taken by the decision maker at stage $t$, respectively.
\item The next state is chosen according to $P(s_{t+1}|s^t,a^t)=P(s_{t+1}|s_t,a_t)$, i.e., the state is Markov conditioned on $a_t$.
\end{itemize}
We define our state as $s_t = P_{X_t|Y^{t-1}}(\cdot|y^{t-1})$ and our actions $a_t : \calX\to\calY$. We note that for every history $x^{t-1}$, the general deterministic encoder (which is a function of $x^t$) is a mapping from $x_t$ to $y_t$. Our action, $a_t$, is this mapping. Since there is only a finite number of mappings from $X_t$ to $Y_t$, our action space is finite. Our state space is also finite. This is true since we consider only deterministic encoders, from which there is only a finite number. Therefore, at each stage, there is only a finite number of possible $P_{X_t|Y^{t-1}}$. This means that the cardinality of the state alphabet, grows with the time horizon $T$. Note however, that the decoder's state alphabet, $\calZ_t=\calY^t$, grows as well in this case.
We start by showing that the cost function can be written as a function of the current state and action.
Treating the codeword length first:
\begin{align}
    L_{Y_t|Y^{t-1}}(y^{t-1}) &= \min_{l(\cdot)\in\calA}\sum_{y_t}P(y_t|y^{t-1})l(y_t)\nt\\
    &= \min_{l(\cdot)\in\calA}\sum_{y_t,x_t}P(y_t,x_t|y^{t-1})l(y_t)\nt\\
    &= \min_{l(\cdot)\in\calA}\sum_{y_t,x_t}P(x_t|y^{t-1})P(y_t|x_t,y^{t-1})l(y_t)\nt\\
    &= \min_{l(\cdot)\in\calA}\sum_{y_t,x_t}P(x_t|y^{t-1})\ind{a_t(x_t)=y_t}l(y_t)\nt\\
    &\eqd \alpha_t(s_t,a_t),
\end{align}
where the equation preceding the last one is true since we know the function from $x_t$ to $y_t$.
We now move on to the average distortion. We first show that the optimal reproduction function, $\hat{X}_{Bayes}(P_{X_t|Y^t})$, is a function of $(a_t,s_t,y_t)$. To see this note that
\begin{align}
    P(x_t|y^t)&=\frac{P(x_t,y_t|y^{t-1})}{\sum_{x_t}P(x_t,y_t|y^{t-1})}\nt\\
    &=\frac{P(x_t|y^{t-1})\ind{a_t(x_t)=y_t}}{\sum_{x_t}P(x_t|y^{t-1})\ind{a_t(x_t)=y_t}}\nt\\
    &\eqd f(s_t,a_t,x_t,y_t).
\end{align}
Therefore, the optimal reproduction function, which is a function of $P_{X_t|Y^t}(\cdot|y^t)$, is a function of $(s_t,a_t,y_t)$, i.e., $\hX_{Bayes}(P_{X_t|Y^t})=g^*_t(s_t,a_t,y_t)$. Using this notation we have
%The average distortion given past encoder outputs can be written as
\begin{align}
    \bE\left[\rho(X_t,g^*_t(s_t,a_t,Y_t))\bigg|Y^{t-1}=y^{t-1}\right]
    &=\sum_{x_t,y_t}P(x_t,y_t|y^{t-1})\rho(x_t,g^*_t(s_t,a_t,y_t))\nt\\
    &=\sum_{x_t,y_t}P(x_t|y^{t-1})P(y_t|x_t,y^{t-1})\rho(x_t,g^*_t(s_t,a_t,y_t))\nt\\
    &=\sum_{x_t,y_t}P(x_t|y^{t-1})\ind{a_t(x_t)=y_t}\rho(x_t,g^*_t(s_t,a_t,y_t))\nt\\
    &\eqd\beta_t(s_t, a_t).
\end{align}
Denoting $\beta_t(s_t, a_t) +\lambda \alpha_t(s_t, a_t) = \gamma_t(s_t, a_t)$, our optimality criterion can be written as $\frT\sum_{t=1}^T \bE\gamma_t(s_t, a_t)$.

We move on to show that the state sequence is Markov conditioned on the action, namely, $P(s_{t+1}|s^{t},a^{t}) = P(s_{t+1}|s_{t},a_t)$.
We start by noting that $s_{t+1}=P_{X_{t+1}|Y^{t}}(\cdot|y^{t})$ is a function of $(a_t,s_t,y_t)$. For every $x_{t+1}$, we have
\begin{align}
    P(x_{t+1}|y^{t}) &= \frac{\sum_{x_t}P(x_{t+1},x_t,y_t|y^{t-1})} {\sum_{x_t,x_{x+1}}P(x_{t+1},x_t,y_t|y^{t-1})}\nt\\
    &= \frac{\sum_{x_t}P(x_t|y^{t-1})P(x_{t+1}|x_t,y^{t-1})P(y_t|x_t,x_{t+1},y^{t-1})} {\sum_{x_t,x_{x+1}}P(x_{t+1},x_t,y_t|y^{t-1})}\nt\\
    &= \frac{\sum_{x_t}P(x_t|y^{t-1})P(x_{t+1}|x_t)\ind{a_t(x_t)=y_t}} {\sum_{x_t,x_{x+1}}P(x_{t+1},x_t,y_t|y^{t-1})}\nt\\
    &= \frac{\sum_{x_t}P(x_t|y^{t-1})P(x_{t+1}|x_t)\ind{a_t(x_t)=y_t}} {\sum_{x_t,x_{x+1}}P(x_t|y^{t-1})P(x_{t+1}|x_t)\ind{a_t(x_t)=y_t}}\nt\\
    &\eqd f(a_t,s_t,x_{t+1},y_t). \label{eq:RecursiveState}
\end{align}
Therefore, $s_{t+1}=h(a_t,s_t,y_t)$, for a function $h$ that uses \eqref{eq:RecursiveState} for every $x_{t+1}$. Now,
\begin{align}
    P(s_{t+1}=\nu|s^t,a^t)&=\sum_{y_t,x_t}\ind{h(a_t,s_t,y_t)=\nu}\ind{a_t(x_t)=y_t}P(x_t|y^{t-1})\nt\\
    &=P(s_{t+1}=\nu|s_t,a_t),
\end{align}
since the current prior on $x_{t}$ is given.
We showed that our system can be represented as an MDP. By invoking Theorem \ref{Thm:AppMDP}, we know that the optimal action at each stage, $a_t$, is a deterministic function of the state. Namely, The mapping from $x_t$ to $y_t$ can be chosen deterministically as a function of $P_{X_t|Y^{t-1}}(\cdot|y^{t-1})$. Therefore, $Y_t$ is a deterministic function of $(X_t,P_{X_t|Y^{t-1}}(\cdot|y^{t-1}))$, which concludes the proof of Theorem \ref{Thm:InfMem}.
Since the state can be recursively calculated (see eq. \eqref{eq:RecursiveState}), the encoder does not need to store $y^{t-1}$ but rather a probability measure (a vector in $\mathbbm{R}^{|\calY|}$).

%Theorem \ref{Thm:InfMem} is more explicit than Theorem \ref{Thm:NoSI} regarding the role of $Z_{t-1}$ at the encoder. It is quite similar to the result of \cite{WalrandVaraiya83} with fixed rate. Reviewing the previous section's proof of the 2--stage and 3--stage lemmas, we see that although the the encoders where shown to be deterministic functions of the current source symbol and decoder state, the optimal encoder at each stage depend on the state of the system ``state'', $P(X_t|Z_{t-1})$, which is determined by the choices of the previous encoders.

%\subsubsection{Constrained rate}
%if the above is correct, we can use literature on constrained MDP's \cite{BeutlerRoss85},\cite{AltmanBook99} to show that in a constrained setting, 2 encoders (one bellow the constraint and one above) need to be used randomly (in the finite horizon case). the encoders are the same except one of the states. This needs more time...
% 

%% file: MarkovMem.tex
\section{Markov Memory Update Functions}\label{Sec:MarkovMemUpdate}
%\subsection*{Introduction}
\subsection{Preliminaries and main result}
In Section \ref{Sec:NoSI}, we showed that for given memory update, distortion and reproduction functions, there is no loss of optimality if the encoders use the current source symbol and the state of the decoder, which they track. We will refer to this class of encoders as \textit{tracking encoders}. In the overall optimization of the system, there is still the task of finding the best memory update and reproduction functions at each stage. When the memory update functions and encoders are fixed, as we discussed in Section \ref{Sec:InfMemNoSI}, the reproduction function should output the $\hX_t$ that minimizes the average distortion for a given $(Y_t,Z_{t-1})$, i.e., the Bayes response of $P_{X_t|Y_t,Z_{t-1}}$. This is simple since the reproduction function has no influence on the future costs (cost to go) and it affects only the present distortion (in a way, for the same reasons, the two--stage lemma was simpler than the three--stage lemma). However, similarly to the encoders, the memory update function at stage $t$ affects all future costs. In this section, we show that for a ``small'' cost at each stage, one can take Markov memory update functions, defined as sliding windows over the received symbols at the decoder and avoid the search for the $|\calZ|$--states optimal memory update functions. The extra cost is a function of $|\calZ|$ and the sliding window size only and it vanishes as the window size is increased.

Let
\begin{align}
    \Delta_{|\calZ|} &= \min_{\{r_t\}\{f_t\}, \{g_t\}}\bE\frT\left\{\sum_{t=1}^T\left[ \rho(X_t, g_t(Y_t, Z_{t-1})) + \lambda L_{Y_t|Z_{t-1}}(Z_{t-1})\right]\right\}\nt\\\label{eq:DeltaZ}
\end{align}
where the minimization is over all next state functions $\{r_t\}$ with a state set of size $|\calZ|$ and all decoders and tracking encoders that use. Note that we choose here the whole sequence of next--state functions, encoders and decoders for $t=1,2,\ldots,T$. We say that the state is Markov of length $l$, if $Z_t=\{Y_{t-l},\ldots,Y_{t-1} \}$, i.e., a sliding window of length $l$ on the encoder outputs.
Let
\begin{align}
    \tilde{\Delta}_{l} = \min_{\{\tilde{f}_t\}, \{\tilde{g}_t\}}\bE\frT\sum_{t=1}^T \left[\rho(X_t, \tilde{g}_t(Y_t, Y_{t-l}^{t-1})) + \lambda L_{Y_t|Y_{t-l}^{t-1}}(Y_{t-l}^{t-1})\right]. \label{eq:Deltal}
\end{align}
where here, the minimization is with respect to all decoders and tracking encoders that use a Markov state of length $l$.

\begin{theorem}\label{Thm:MarkovMem}
For any source statistics, when considering only tracking encoders, we have for any $l$ that divides $T$:
\begin{align}
    \Delta_{\calZ} \geq \tilde{\Delta}_l - \lambda\frac {\log|\calZ|} l
\end{align}
\end{theorem}
The significance of this theorem is more conceptual than operational. The system on the r.h.s might require more memory than the system on the l.h.s. and the search for the optimal encoders becomes more complex as $l$ increases. However, the system on the r.h.s is conceptually simpler and analytically more tractable since the memory structure is simple.

Combining Theorem \ref{Thm:MarkovMem} with Theorem \ref{Thm:NoSI} we have the following theorem:
\begin{theorem}\label{Thm:Combined}
    For a Markov source, there exists a system with deterministic encoders $Y_t=f_t(X_t,Z_{t-1})$ and Markov memory update functions with a performance loss no greater than $\lambda\frac {\log|\calZ|} l$ per source symbol, compared to the optimal system.
\end{theorem}

Theorem \ref{Thm:MarkovMem} can be extended to the case where instead of our Lagrangian cost function, we would look for the minimal average distortion subject to an average length constraint. Let
\begin{align}
    \Delta_{\calZ}(R) &\eqd \min_{\{f_t\},\{g_t\},\{r_t\}} \bE\cbr{\frT\sum_{t=1}^T\rho(X_t,g_t(Y_t,Z_{t-1}))}\nt\\
    &~~~~~~~~~~~    s.t ~~~~~~~ \bE\cbr{\frT\sum_{t=1}^T L_{Y_t}(Z_{t-1})}<R\nt\\
    \tilde{\Delta}_{l}(R) &\eqd \min_{\{f_t\},\{g_t\}} \bE\cbr{\frT\sum_{t=1}^T\rho(X_t,g_t(Y_{t-l}^t))}\nt\\
    &~~~~~~~~~~~    s.t  ~~~~~~~ \bE\cbr{\frT\sum_{t=1}^T L_{Y_t}(Y_{t-l}^{t-1})}<R
\end{align}
where the minimization is over all tracking encoders that use $X_t$ and the decoder's state, reproduction functions and state update functions (in $\Delta_{\calZ}(R)$ only).
We have the following theorem:
\begin{theorem} \label{Thm:ConstrainedRate}
In the constrained setting, for any $l$ that divides $T$ we have
\begin{align}
    \Delta_{\calZ}(R) \geq \tilde{\Delta}_l\lb R+\frac{\log|\calZ|} l\rb
\end{align}
\end{theorem}
Note that here we do not have a theorem in the spirit of Theorem \ref{Thm:Combined} since we did not show that in this case, tracking encoders are optimal.

In the next subsection, we prove Theorem \ref{Thm:MarkovMem}. Theorem \ref{Thm:Combined} is a direct consequence of Theorem \ref{Thm:NoSI} and Theorem \ref{Thm:MarkovMem} combined. Theorem \ref{Thm:ConstrainedRate} is proven exactly in the same manner as Theorem \ref{Thm:MarkovMem} and its proof is therefore, omitted. Theorem \ref{Thm:MarkovMem} is valid even without taking expectations in \eqref{eq:DeltaZ},\eqref{eq:Deltal} and therefore, it is also valid for individual sequences (see \cite{NeriZiv06}). Theorems \ref{Thm:MarkovMem}--\ref{Thm:ConstrainedRate} will also hold in the setting of the Section \ref{Sec:SI}, where SI is available to the decoder.

\subsection{Proof of Theorem \ref{Thm:MarkovMem}:}
The ideas in the proof rely on some ideas from \cite{NeriZiv06}. Fix the optimal encoders, state update and reproduction functions of $\Delta_{|\calZ|}$.
We start by focusing on the codeword length element of $\Delta_{|\calZ|}$, using the fact that conditioning reduces the length element (see Appendix \ref{App:LenProperties}), we have
\begin{align}
    R &\eqd\frT\sum_{t=1}^T \bE L_{Y_t|Z_{t-1}}(Z_{t-1})\nt\\
    &\geq \frT\sum_{t=1}^T \bE L_{Y_t|Y_{t-l}^{t-1},Z_{t-1}}(Y_{t-l}^{t-1},Z_{t-1})\label{eq:MarkovMemStart}
\end{align}
Since we will always deal with the expected codeword length, in order to simplify the notation, we will use from now $\bE L_{Y_t|Z_{t-1}}(Z_{t-1}) \eqd L_{Y_t|Z_{t-1}}$ (as defined in Section \ref{Sec:Prelim}).
We now add conditioning on $Z_0, Z_l, Z_{2l},\ldots$ which will further reduce the last expression. $Z_0$ is added to the first $l$ summands of \eqref{eq:MarkovMemStart}, $Z_l$ to the summands indexed by $l+1,\ldots,2l$, and so on. This conditioning makes the conditioning on $Z_{t-1}$ redundant since if we know the state in the past and the encoder outputs up to the present, we know the current state as well. We continue by assuming that $l$ divides $T$:
\begin{align}
    R &\geq \frT\sum_{t=1}^T L_{Y_t|Y_{t-l}^{t-1},Z_{t-1}}\nt\\
      &\geq \frT\sum_{j=0}^{T/l-1}\sum_{t=jl+1}^{jl+l} L_{Y_t|Y_{t-l}^{t-1},Z_{jl}}\label{dsum}
\end{align}
Now there are two types of terms:
\begin{enumerate}
    \item $(Y_{jl+1},Z_{jl})$ appear together in the conditioning.
    \item $Y_{jl+1}$ is conditioned on $Z_{jl}$ and the previous block: $Y_{j(l-1)+1}^{jl}$.
\end{enumerate}
We rewrite the sum of \eqref{dsum} as two sums, pertaining to the above two types:
\begin{align}
    R &\geq \frT\sum_{j=0}^{T/l-1}\sum_{t=jl+1}^{jl+l} L_{Y_t|Y_{t-l}^{t-1},Z_{jl}}\nt\\
        &= \frT\sum_{j=0}^{T/l-1}\sum_{t=jl+2}^{jl+l}  L_{Y_t|Y_{t-l}^{t-1},Z_{jl}}+ \frT\sum_{j=0}^{T/l-1} L_{Y_{jl+1}|Y_{j(l-1)+1}^{jl},Z_{jl}}.\label{eq:sums}
\end{align}
We now use the following inequality, which is proved in Appendix \ref{App:LenProperties}:
\begin{align}
    L_{Y_{jl+1}|Y_{j(l-1)+1}^{jl},Z_{jl}} \geq L_{Y_{jl+1},Z_{jl}|Y_{j(l-1)+1}^{jl}} -\log |\calZ| \label{eq:simple}
\end{align}
Substituting \eqref{eq:simple} in \eqref{eq:sums}, we have:
\begin{align}
    R &\geq \frT\sum_{j=0}^{T/l-1}\sum_{t=jl+2}^{jl+l} L_{Y_t|Y_{t-l}^{t-1},Z_{jl}}+ \frT\sum_{j=0}^{T/l-1} L_{Y_{jl+1}|Y_{j(l-1)+1}^{jl},Z_{jl}}\nt\\
    &\geq \frT\sum_{j=0}^{T/l-1}\sum_{t=jl+2}^{jl+l}  L_{Y_t|Y_{t-l}^{t-1},Z_{jl}}+ \frT\sum_{j=0}^{T/l-1} L_{Y_{jl+1},Z_{jl}|Y_{j(l-1)+1}^{jl}} - \frac {\log |\calZ|} l\nt\\
    &\geq \frT\sum_{j=0}^{T/l-1}\sum_{t=jl+2}^{jl+l}  L_{Y_t|Y_{t-l}^{t-1},Z_{jl}}+ \frT\sum_{j=0}^{T/l-1} L_{Y_{jl+1},Z_{jl}|Y_{j(l-1)+1}^{jl},Z_{j(l-1)}} - \frac {\log |\calZ|} l.\label{eq:Length}
\end{align}

Regarding the distortion element of $\Delta_{|\calZ|}$, we have:
\begin{align}
    &\frT \sum_{t=1}^T \bE\rho(X_t, g_t(Y_t, Z_{t-1})) \nt\\
    &~~= \frT\sum_{j=0}^{T/l-1}\sum_{t=jl+2}^{jl+l} \bE\rho(X_t, g_t(Y_t, Z_{t-1})) + \frT\sum_{j=0}^{T/l-1} \bE\rho(X_{jl+1},g_t(Y_{jl+1}, Z_{jl}))\nt\\
    &~~\geq \frT\min_{\{g_t\}}\left\{\sum_{j=0}^{T/l-1}\sum_{t=jl+2}^{jl+l} \bE\rho(X_t, g_t(Y_{t-l}^{t}, Z_{jl}))
    + \sum_{j=0}^{T/l-1} \bE\rho(X_{jl+1},g(Y_{j(l-1)+1}^{jl+1},Z_{jl},Z_{j(l-1)}))\right\}.\label{eq:dist}
\end{align}
In the last inequality, we used the fact that with the same encoders (same $\{Y_t\}$), optimal decoders that use more data will do at least as well as the original decoders.
%Now, $Y_t$ is a function of $X_t$ and $Z_{t-1}$. For every $t=jl+2,\ldots,jl+l~~~~~~ j=0,1,\ldots, n/l-1$, $Y_t$ is therefore a different function of $X_t, Z_{jl}, Y_{t-l}^t-1$ (which is essentially calculates $Z_{t-1}$ and than, using $X_t$, outputs the same $Y_t$ as the original function). Also, $(Y_{jl+1},Z_{jl})$ is in the same manner a function of $X_t, Y_{j(l-1)+1}^{jl},Z_{j(l-1)}$.
Note that in the above derivation, $(Y_{jl+1}, Z_{jl})$ always appear together. Therefore, we set for all $j=0,1,\ldots, n/l-1$ $Y_{jm+1}' = (Y_{jm+1},Z_{jl})$ and for all other indexes we set $Y_t'=Y_t$. Using this notation, we have for \eqref{eq:Length}:
\begin{align}
    R \geq \frT\sum_{t=1}^n L_{Y_t'}(Y_{t-l}^{'t-1}) - \frac {\log |\calZ|} l.
\end{align}
and for \eqref{eq:dist}
\begin{align}
    \frT \sum_{t=1}^T \bE\rho(X_t, g_t(Y_t, Z_{t-1})) &\geq \min_{\{g_t\}}\frT\sum_{t=1}^T\bE \rho(X_t, g_t(Y_{t-l}^{'t})).
\end{align}
and each $Y'_t$ is a function of $X_t, Y_{t-l}^{'t-1}$.
Note that although the size of the alphabet of $Y'$ is now $|\calY|\times|\calZ|$, the size of the alphabet was not a constraint on the system and was introduced so it will be convenient to define $\calA$. The fact that it is now larger does not change any of the results obtained in the previous sections.
We have
\begin{align}
    \Delta_{\calZ} \geq \min_{\{g_t\}}\frT\sum_{t=1}^T \bE\rho(X_t, g_t(Y_{t-l}^{'t})) + \lambda\lb\frT\sum_{t=1}^n \bE L_{Y_t'|Y_{t-l}^{'t-1}}(Y_{t-l}^{'t-1}) - \frac {\log |\calZ|} l\rb.
\end{align}
The r.h.s of the above equation was calculated with the optimal encoders of the l.h.s. with a scheme that appends the original decoder state once every block. This is, of course, only one of the possible schemes for Markovian states and therefore if we optimize the r.h.s over all encoders that use a Markovian state of length $l$ we get
\begin{align}
    \Delta_{\calZ} \geq \tilde{\Delta}_l - \lambda\frac {\log|\calZ|} l
\end{align}
\IEEEQED 

%% file: Appendix.tex
\appendix
\setcounter{equation}{0}
\setcounter{theorem}{0}
\numberwithin{equation}{section}
\numberwithin{theorem}{section}
%\subsection{The set of stochastic encoders is a convex set}\label{App:ConvexSet}
\subsection{Proof of Lemma \ref{Lem:2StageConcave}} \label{App:2ndCostConcave}
We start by focusing on the average codeword length element of the cost function and show that $L_{Y_t}(Z_{t-1})$ is concave in $\{f_{X_t,Z_{t-1}}^s\}$. For $0\leq\alpha\leq 1$ and $f_1,f_2\in\{f_{X_t,Z_{t-1}}^s\}$, let
\begin{align*}
    f_{\alpha}= \alpha f_1+(1-\alpha)f_2.
\end{align*}
This means that for any $(x_t,z_{t-1})$ we have
\begin{align}
    P_{\alpha}(y_t|x_t,z_{t-1}) &= \alpha P_{1}(y_t|x_t,z_{t-1}) + (1-\alpha)P_{2}(y_t|x_t,z_{t-1})
\end{align}
Let $L^{f_{\alpha}}_{Y_t}(Z_{t-1})$, $L^{f_1}_{Y_t}(Z_{t-1})$, $L^{f_2}_{Y_t}(Z_{t-1})$ denote the length function calculated with $f_{\alpha}$, $f_1$, $f_2$ respectively.
We have
\begin{align}
    &L^{f_{\alpha}}_{Y_t}(z_{t-1}^y)=\min_{l(\cdot)\in\calA}\bigg\{\sum_{y_t}P_{\alpha}(y_t|z_{t-1})l(y_t) \bigg\}\nt\\
    &=\min_{l(\cdot)\in\calA}\bigg\{\sum_{y_t,x_t} \left[\alpha P_{1}(y_t|x_t,z_{t-1})\right.\nt\\
    &+\left.(1-\alpha)P_{2}(y_t|x_t,z_{t-1})\right]P(x_t|z_{t-1})l(y_t) \bigg\}\nt\\
    &\geq \alpha\min_{l(\cdot)\in\calA}\bigg\{\sum_{y_t,x_t} P_{1}(y_t|x_t,z_{t-1})P(x_t|z_{t-1})l(y_t)\bigg\}
    +(1-\alpha)\min_{l(\cdot)\in\calA}\bigg\{\sum_{y_t,x_t} P_{2}(y_t|x_t,z_{t-1})P(x_t|z_{t-1})l(y_t)\bigg\}\nt\\
    &=\alpha L^{f_1}_{Y_t}(z_{t-1})+(1-\alpha)L^{f_2}_{Y_t}(z_{t-1})
\end{align}
where we used the fact that the sum of minima is smaller than the minimum of a sum. Since the distortion part of the cost is linear in $P(y_t|x_t,z_{t-1})$ (through the expectation), we have that the overall stage $t$ cost function is concave in $\{f_{X_t,Z_{t-1}}^s\}$.

\subsection{Proof of Lemma \ref{Lem:3StageConcave}}\label{App:3StageConcave}
Fix any third stage encoder which is a deterministic function of $(X_3,Z_2)$.
We showed in \eqref{eq:D3} that the second stage encoder affects $J_3$ only through $P(x_3,z_2)$ (and thus also through $P(z_2)$ and $P(x_3|z_2)$). Let $f_1,f_2\in\{f_{X_2Z_1}^s\}$ be two second stage stochastic encoders which are functions of $(X_2,Z_1)$.
Let
%\subsection{The third stage cost is concave in $\{P(Y_2|X_2,Z_1) \}$} \label{App:3rdCostConcave}
%in this Appendix, we prove lemma \ref{Lem:3StageConcave}.
%
%\noindent {\it Proof}: Fix any third stage encoder. We write the third stage cost:
%\begin{align}
%    J_3 &= \bE\left\{\rho(X_3, g(Y_3,Z_2))+\lambda L_{Y_3|Z_2}(Z_2) \right\}\nt\\
%    &= \sum_{x_3,y_3,z_2}P(x_3,z_2)P(y_3|x_3,z_2)\left\{  \rho(x_3, g(y_3,z_2))+\lambda L_{Y_3|Z_2}(z_2) \right\}\nt\\
%    &= \sum _{x_3,y_3,z_2}P(x_3,z_2)P(y_3|x_3,z_2) \rho(x_3, g(y_3,z_2)) + \lambda \sum_{z_2}P(z_2)\min_{l(\cdot)\in\calA}\sum_{y_3}P(y_3|z_2)l(y_3).\label{eq:D3}
%\end{align}
%The second stage encoder, affects the last expression through $P(X_3,Z_2)$ (and thus also through $P(Z_2)$ and $P(X_3|Z_2)$) since
%\begin{align}
%    P(x_3,z_2) &= \sum_{x_2,y_2,z_1}P(x_2,x_3,y_2,z_1,z_1)\nt\\
%    &= \sum_{x_2,y_2,z_1} P(x_2,z_1)P(y_2|x_2,z_1)P(z_2|z_1,y_2)P(x_3|x_2)\label{eq:PX3Z2}
%\end{align}
%where $P(x_2,z_1)$ is the result of the first stage and we used the fact that the source is Markov and that $z_2$ is a deterministic function of $y_2,z_1$.\\
%Let
\begin{align}
    P_{\gamma}(x_3,z_2) &= \sum_{x_2,y_2,z_1} P(x_2,z_1)\sbr{\gamma P_1(y_2|x_2,z_1)+(1-\gamma) P_2(y_2|x_2,z_1)}P(z_2|z_1,y_2)P(x_3|x_2),\nt\\
    &= \gamma P_{1}(x_3,z_2) + (1-\gamma) P_{2}(x_3,z_2),\label{eq:P_lambda}
\end{align}
where $P_{1}(x_3,z_2), P_{2}(x_3,z_2)$ are calculated with $P_1(y_2|x_2,z_1), P_2(y_2|x_2,z_1)\}$ that result from $f_1,f_2$ respectively. Similarly, for $i=1,2,\gamma$, define $P_i(z_2)$, and  $P_i(x_3|z_2)$ as the marginal and conditional distribution, respectively, resulting from the probability measures in \eqref{eq:P_lambda}.
We now show that $P_{\gamma}(x_3|z_2)$ can be written as a convex combination of $P_1(x_3|z_2), P_2(x_3|z_2)$.
\begin{align}
P_{\gamma}(x_3|z_2) &= \frac{P_{\gamma}(x_3,z_2)} {P_{\gamma}(z_2)}\nt\\
&=\frac{\gamma P_{1}(x_3,z_2) + (1-\gamma)P_{2}(x_3,z_2)} {\sum_{x'_3} \gamma P_{1}(x'_3,z_2) + (1-\gamma)P_{2}(x'_3,z_2)}\nt\\
&= \frac{\gamma P_{1}(x_3,z_2)} {\sum_{x'_3} \gamma P_{1}(x'_3,z_2) + (1-\gamma)P_{2}(x'_3,z_2)} + \frac{(1-\gamma)P_{2}(x_3,z_2)} {\sum_{x'_3} \gamma P_{1}(x'_3,z_2) + (1-\gamma)P_{2}(x'_3,z_2)}\nt\\
&=\alpha \frac{P_{1}(x_3,z_2)} {\sum_{x'_3} P_{1}(x'_3,z_2)} + \beta \frac{P_{2}(x_3,z_2)} {\sum_{x'_3} P_{2}(x'_3,z_2)}
\end{align}
with,
\begin{align}
    \alpha &= \frac{\gamma \sum _{x'_3} P_{1}(x'_3,z_2)} {\sum_{x'_3} \gamma P_{1}(x'_3,z_2) + (1-\gamma)P_{2}(x'_3,z_2)}=\frac{\gamma P_1(z_2)} {P_{\gamma}(z_2)}\nt\\
    \beta &= \frac{(1-\gamma)\sum_{x'_3}P_{2}(x'_3,z_2)} {\sum_{x'_3} \gamma P_{1}(x'_3,z_2) + (1-\gamma)P_{2}(x'_3,z_2)}=\frac{(1-\gamma) P_2(z_2)} {P_{\gamma}(z_2)}.\label{eq:alphabeta1}
\end{align}
Note that $0\leq \alpha,\beta\leq 1$ and $\alpha+\beta=1$. We showed that
\begin{align}
    P_{\gamma}(x_3|z_2) = \alpha P_1(x_3|z_2) + (1-\alpha)P_2(x_3|z_2) \label{eq:alphabeta}
\end{align}

We are now ready to prove the lemma.
For any given third stage encoder, let $J_3(P_i)$, $i=1,2,\gamma$, denote the third stage cost as a function of the joint probability of $(X_3,Z_2)$, where the dependence on the second stage encoder was shown in \eqref{eq:PX3Z2},\eqref{eq:P_lambda}. In order to prove the lemma, we need to show that:
\begin{align}
    &J_3(P_{\gamma}) \geq \gamma J_3(P_1) +(1-\gamma)J_3(P_2).
\end{align}

We now focus on the codeword length element of the cost function. Let $L_3(P_i)$, $i=1,2,\gamma$, denote the third stage average codeword length as a function of the joint probability of $(X_3,Z_2)$
\begin{align}
    L_3(P_{\gamma}) &= \sum_{z_2}P_{\gamma}(z_2)\min_{l(\cdot)\in\calA}\sum_{y_3}P_{\gamma}(y_3|z_2)l(y_3)\nt\\
    &= \sum_{z_2}P_{\gamma}(z_2)\min_{l(\cdot)\in\calA}\sum_{y_3,x_3}P_{\gamma}(y_3,x_3|z_2)l(y_3)\nt\\
    &= \sum_{z_2}P_{\gamma}(z_2)\min_{l(\cdot)\in\calA}\sum_{y_3,x_3}P_{\gamma}(x_3|z_2)P(y_3|x_3,z_2)l(y_3)\nt\\
    &= \sum_{z_2}P_{\gamma}(z_2)\min_{l(\cdot)\in\calA}\sum_{y_3,x_3}\left[\alpha P_{1}(x_3|z_2) + (1-\alpha)P_{2}(x_3|z_2)\right]P(y_3|x_3,z_2)l(y_3)\label{eq:LConcave1}\\
    &\geq \sum_{z_2}P_{\gamma}(z_2)\alpha\min_{l(\cdot)\in\calA}\sum_{y_3,x_3} P_{1}(x_3|z_2)P(y_3|x_3,z_2)l(y_3)+\nt\\
    &~~~~~~~~\sum_{z_2}P_{\gamma}(z_2)(1-\alpha)\min_{l(\cdot)\in\calA}\sum_{y_3,x_3} P_{2}(x_3|z_2)P(y_3|x_3,z_2)l(y_3)\label{eq:LConcave2}\\
    &=\sum_{z_2}P_{\gamma}(z_2)\frac{\gamma P_1(z_2)} {P_{\gamma}(z_2)} \min_{l(\cdot)\in\calA}\sum_{y_3,x_3} P_{1}(x_3|z_2)P(y_3|x_3,z_2)l(y_3)+\nt\\
    &~~~~~~~~\sum_{z_2}P_{\gamma}(z_2)\frac{(1-\gamma) P_2(z_2)} {P_{\gamma}(z_2)}\min_{l(\cdot)\in\calA}\sum_{y_3,x_3} P_{2}(x_3|z_2)P(y_3|x_3,z_2)l(y_3)\label{eq:LConcave3}\\
    &=\gamma d_3(P_{1})+(1-\gamma)d_3(P_2),\label{eq:LConcave4}
\end{align}
where in \eqref{eq:LConcave1} we used \eqref{eq:alphabeta}, \eqref{eq:LConcave2} is true since the minimum of a sum is greater than the sum of minima and finally, in \eqref{eq:LConcave3} we substituted $\alpha$ given in \eqref{eq:alphabeta1}.
Thus, we showed that the codeword length element of the cost function is concave in the choice of the second stage encoder. We have
\begin{align}
    J_3(P_{\gamma}) &= \sum_{x_3,y_3,z_2} P(y_3|x_3,z_2)P_{\gamma}(x_3,z_2)\rho(x_3, g(y_3,z_2)) + \lambda L_3(P_{\gamma})\nt\\
    &= \sum_{x_3,y_3,z_2} \gamma P(y_3|x_3,z_2)P_{1}(x_3,z_2)\rho(x_3, g(y_3,z_2)) + \nt\\
     &~~~~~~(1-\gamma) P(y_3|x_3,z_2)P_{2}(x_3,z_2)\rho(x_3, g(y_3,z_2))+ \lambda L_3(P_{\gamma})\nt\\
    &\geq \gamma J_3(P_{1}) + (1-\gamma)J_3(P_{2})
\end{align}
where in the last step we used \eqref{eq:LConcave4} and the lemma is proven.
\IEEEQED

\subsection{Markov decision processes - short overview} \label{App:MDPs}
In a Markov decision process, a decision maker is influencing the behavior of a Markov probabilistic system through his actions, as the system evolves in time.
Formally, a discrete time, finite horizon Markov decision process is defined by $\{T,S,A, \{P_t(\cdot|s,a)\}, \{\rho_t(s,a)\} \}$, where,
\begin{itemize}
    \item $T$ is the time horizon, $t=1,2,\ldots,T$.
    \item $\calS$ is the state space.
    \item $\calA$ is the action space.
    \item $P_t(\cdot|s,a)$ is the transition probability to the systems's next state, given the previous system state and action. The transition probabilities obey $P_t(\cdot|s^t,a^t)=P_t(\cdot|s_t,a_t)$, namely, the next state, $s_{t+1}$, distribution depends on the history only through $(s_t, a_t)$.
    \item $P_0(\cdot)$ is the probability measure over the initial state.
    \item $\rho_t(s,a)$ is the cost incurred when at stage $t$ and state $s$, action $a$ is taken.
\end{itemize}
In our case, the goal of the decision maker is to minimize the expected average cost $\bE\frT\sum_{t=1}^T\rho_t(S_t,A_t)$.
The history of the process at stage $t$ is $h_t=(s_1,a_1,s_2,a_2,\ldots,s_{t-1},a_{t-1},s_t)$, i.e., all previous actions taken by the decision maker and the system states, up to stage $t$. Note that $h_t=\{h_{t-1},a_{t-1},s_t\}$.

A decision rule, $d_t$, prescribes the procedure for action selection in a given state at stage $t$. Decision rules can range from deterministic functions of the current state to randomized functions that depend on the whole history of states and actions, up to stage $t$. A decision rule that is a deterministic function of the current state will be called a Markovian deterministic (MD) decision rule. A policy specifies the decision rules to be used at all stages, i.e., a policy $\pi$ is a sequence of decision rules $d_1,\ldots,d_T$. We say that a policy is MD if all its decision rules are MD.

In Sections \ref{Sec:InfMemNoSI},\ref{Sec:InfMemSI}, the state space is finite, however, it grows as the system evolves, i.e., at each stage the stage space is $\calS_t$. We set $\calS = \cup_{t=1}^T\calS_t$. The action space is the set of deterministic functions $f: \calX\to\calY$, which is finite.

We will use the following theorem which is the key to the results of section \ref{Sec:InfMemNoSI}.

\begin{theorem}(\cite[Proposition 4.4.3]{PutermanBook94}):\label{Thm:AppMDP}
    There exist an MD policy which is optimal.
\end{theorem}
We outline the proof here for completeness.\\
\textit{Proof of Theorem \ref{Thm:AppMDP} (outline):}
Define for policy $\pi$, $u^{\pi}_t(h_t) = \bE\left\{\sum_{i=t}^T \rho_i(s_i,a_i) | h_t\right\}$, where the actions $a_i$ are prescribed by the policy $\pi$. Note that
\begin{align}
    u^{\pi}_t(h_t)=\rho_t(s_t,a_t)+\sum_{j\in\calS}p_t(j|s_t,a_t)u_{t+1}^{\pi}(h_t,j,a_t)
\end{align}
Let $u^{*}_t(h_t)=\inf_{\pi}u^{\pi}_t(h_t)$. We start by showing the $u_t^{*}(h_t)$ depends on the history only through $s_t$. We will use backwards induction. Note that $u^*_T(h_T)=\min_{a\in\calA}\rho(s_T,a)$, so the claim is valid for the last stage. Now assume that the claim is valid for $n=t+1,t+2,\ldots,T$. We have
\begin{align}
    u^{*}_t(h_t)&=\min_{a\in\calA}\left\{\rho_t(s_t,a)+\sum_{j\in\calS}p_t(j|s_t,a)u_{t+1}^{*}(h_t,j,a)\right\}\nt\\
    &=\min_{a\in\calA}\left\{\rho_t(s_t,a)+\sum_{j\in\calS}p_t(j|s_t,a)u_{t+1}^{*}(j)\right\} \label{eq:appMDP}
\end{align}
where the last equation is due to the induction hypothesis. Since the term in brackets depends on the history only through $s_t$, the induction step is proven. Now, define the decision rule at each stage for every $s_t\in\calS$ as the minimizer of \eqref{eq:appMDP}. By construction, this decision rule is MD and the policy constructed from these decision rules is optimal.

\subsection{Properties of the length function}\label{App:LenProperties}
Let $\calW,\calY,\calZ$ be finite alphabets.

\subsubsection{Conditioning reduces the length}

We have
\begin{align}
    L_{Y|Z}&=\sum_{z\in\calZ} P(z)\min_{l(\cdot)\in\calA}\sum_{y\in\calY}P(y|z)l(y)\nt\\
    &=\sum_{z\in\calZ} P(z)\min_{l(\cdot)\in\calA}\sum_{y\in\calY}\sum_{w\in\calW}P(y|w,z)P(w|z)l(y)\nt\\
    &\geq\sum_{z\in\calZ}\sum_{w\in\calW} P(z)P(w|z)\min_{l(\cdot)\in\calA}\sum_{y\in\calY}P(y|w,z)l(y)\nt\\
    &=L_{Y|W,Z}
\end{align}
where the inequality is true since the minimum of a sum is greater than the sum of minima.

\subsubsection{Proving that $L_{Y|W,Z} \geq L_{Y,Z|W} -\log |\calZ|$}

The intuition behind this is simple: given $W$, the average optimal code length for the pair $(Y,Z)$ can not be larger than coding $Z$ separately and concatenating a codeword that describes $Y$ and is decodable when $Z$ is known. The optimal scheme for coding the pair can not be worse, otherwise, this scheme can be used. To see this mathematically, for each $y\in\calY, z\in\calZ$, let $l^*(y), l^*(z)$ be the length functions optimized for the distributions $P(y|w,z)$ and $P(z|w)$ respectively. Using the fact that $L_{Z|W}<\log|\calZ|$ we have:
\begin{align}
    L_{Y|Z,W} + \log |\calZ| &\geq \sum_{w,z}P(w,z)\sum_{y}P(y|w,z)l^*(y)+ \sum_w P(w)\sum_z P(z|w)l^*(z)\nt\\
    &= \sum_w P(w) \sum_z P(z|w) \left[\left(\sum_y P(y|w,z)l^*(y)\right)  + l^*(z)\right]\nt\\
    &= \sum_w P(w) \sum_z P(z|w)\sum_y P(y|w,z)[l^*(y)  + l^*(z)]\nt\\
    &\geq \sum_w P(w)\min_{l\in\tilde{\calA}}\sum_{z,y} P(y,z|w)l(y,z)\nt\\
    &=L_{Y,Z|W}
\end{align}
where $\tilde{\calA}$ is defined as in Section \ref{Sec:Prelim} with $\calY\times\calZ$ replacing $\calY$.

%% file: Strucutre-Thms.bbl
% Generated by IEEEtran.bst, version: 1.13 (2008/09/30)
\begin{thebibliography}{10}
\providecommand{\url}[1]{#1}
\csname url@samestyle\endcsname
\providecommand{\newblock}{\relax}
\providecommand{\bibinfo}[2]{#2}
\providecommand{\BIBentrySTDinterwordspacing}{\spaceskip=0pt\relax}
\providecommand{\BIBentryALTinterwordstretchfactor}{4}
\providecommand{\BIBentryALTinterwordspacing}{\spaceskip=\fontdimen2\font plus
\BIBentryALTinterwordstretchfactor\fontdimen3\font minus
  \fontdimen4\font\relax}
\providecommand{\BIBforeignlanguage}[2]{{%
\expandafter\ifx\csname l@#1\endcsname\relax
\typeout{** WARNING: IEEEtran.bst: No hyphenation pattern has been}%
\typeout{** loaded for the language `#1'. Using the pattern for}%
\typeout{** the default language instead.}%
\else
\language=\csname l@#1\endcsname
\fi
#2}}
\providecommand{\BIBdecl}{\relax}
\BIBdecl

\bibitem{Witsenhausen1979}
H.~S. Witsenhausen, ``The structure of real time source coders,'' \emph{Bell
  Systems Technical Journal}, vol.~58, no.~6, pp. 1338--1451, July 1979.

\bibitem{WalrandVaraiya83}
J.~C. Walrand and P.~Varaiya, ``Optimal causal coding--decoding problems,''
  \emph{IEEE Transactions on Information Theory}, vol.~29, no.~6, pp. 814--820,
  November 1983.

\bibitem{Teneketzis2006}
D.~Teneketzis, ``On the structure of optimal real-time encoders and decoders in
  noisy communication,'' \emph{IEEE Transactions on Information Theory},
  vol.~52, no.~9, pp. 4017--4035, September 2006.

\bibitem{WynerZiv76}
A.~D. Wyner and J.~Ziv, ``The rate--distortion function for source coding with
  side information at the decoder,'' \emph{IEEE Transactions on Information
  Theory}, vol.~22, no.~1, pp. 1--10, January 1976.

\bibitem{BorkarMitterTatikonda2001}
\BIBentryALTinterwordspacing
V.~S. Borkar, S.~K. Mitter, and S.~Tatikonda, ``Optimal sequential vector
  quantization of {M}arkov sources,'' \emph{SIAM Journal on Control and
  Optimization}, vol.~40, no.~1, pp. 135--148, 2001. [Online]. Available:
  \url{http://link.aip.org/link/?SJC/40/135/1}
\BIBentrySTDinterwordspacing

\bibitem{GaarderSlepian1982}
N.~T. Gaarder and D.~Slepian, ``On optimal finite-state digital transmission
  systems,'' \emph{IEEE Transactions on Information Theory}, vol.~28, no.~3,
  pp. 167--186, March 1982.

\bibitem{GorantlaColeman2011}
S.~K. Gorantla and T.~P. Coleman, ``Information-theoretic viewpoints on optimal
  causal coding-decoding problems,'' \emph{CoRR}, vol. abs/1102.0250, 2011.

\bibitem{AsnaniWeissmann11}
H.~Asnani and T.~Weissman, ``On real time coding with limited lookahead,''
  \emph{CoRR}, vol. abs/1105.5755, 2011.

\bibitem{MahajanTeneketzis2009}
A.~Mahajan and D.~Teneketzis, ``Optimal design of sequential real-time
  communication systems,'' \emph{IEEE Transactions on Information Theory},
  vol.~55, no.~11, pp. 5317--5338, November 2009.

\bibitem{NeuhoffGilbert1982}
D.~Neuhoff and R.~K. Gilbert, ``Causal source codes,'' \emph{IEEE Transactions
  on Information Theory}, vol.~28, no.~5, pp. 701--713, September 1982.

\bibitem{TsachyNeri2005}
T.~Weissman and N.~Merhav, ``On causal source codes with side information,''
  \emph{IEEE Transactions on Information Theory}, vol.~51, no.~11, pp.
  4003--4013, November 2005.

\bibitem{NeriIoannis2003}
N.~Merhav and I.~Kontoyiannis, ``Source coding exponents for zero-delay coding
  with finite memory,'' \emph{IEEE Transactions on Information Theory},
  vol.~49, no.~3, pp. 609--625, March 2003.

\bibitem{NeriZiv06}
N.~Merhav and J.~Ziv, ``On the {W}yner-–{Z}iv problem for individual
  sequences,'' \emph{IEEE Transactions on Information Theory}, vol.~52, no.~3,
  pp. 867--873, March 2006.

\bibitem{AlonOrlitski96}
N.~Alon and A.~Orlitsky, ``Source coding and graph entropies,'' \emph{IEEE
  Transactions on Information Theory}, vol.~42, no.~5, pp. 1329--1339,
  September 1996.

\bibitem{AltmanBook99}
E.~Altman, \emph{Constrained {M}arkov Decision Processes}, ser. Stochastic
  Modeling Series.\hskip 1em plus 0.5em minus 0.4em\relax Chapman and Hall/CRC,
  1999.

\bibitem{PutermanBook94}
M.~Puterman, \emph{Markov decision processes: discrete stochastic dynamic
  programming}, ser. Wiley series in probability and statistics.\hskip 1em plus
  0.5em minus 0.4em\relax Wiley-Interscience, 1994.

\end{thebibliography}
